\documentclass[superscriptaddress,twocolumn]{revtex4}
\usepackage{amsmath,lscape,sidecap,epsfig}

\def\beq{\begin{equation}}
\def\eeq{\end{equation}}
\def\beqa{\begin{eqnarray}}
\def\eeqa{\end{eqnarray}}
\def\ban{\begin{eqnarray*}}
\def\ean{\end{eqnarray*}}
\def\bi{\begin{itemize}}
\def\ei{\end{itemize}}

\begin{document}

\title{$\alpha$ particles and the pasta phase} 

\author{S. S. Avancini}
\affiliation{Depto de F\'{\i}sica - CFM - Universidade Federal de Santa
Catarina  Florian\'opolis - SC - CP. 476 - CEP 88.040 - 900 - Brazil}
\author{C. C. Barros Jr.}
\affiliation{Depto de F\'{\i}sica - CFM - Universidade Federal de Santa
Catarina  Florian\'opolis - SC - CP. 476 - CEP 88.040 - 900 - Brazil}
\author{D.P.Menezes}
\affiliation{Depto de F\'{\i}sica - CFM - Universidade Federal de Santa
Catarina  Florian\'opolis - SC - CP. 476 - CEP 88.040 - 900 - Brazil}
\author{C. Provid\^encia}
\affiliation{Centro de F\'{\i}sica Computacional, Department of Physics,
University of Coimbra, P 3004 - 516,  Coimbra, Portugal}

\begin{abstract}
 The effects of the $\alpha$ particles in nuclear matter at low densities
are investigated within three different parametrizations of relativistic models
at finite temperature. Both homogeneous and inhomogeneous matter (pasta phase) 
are described for neutral nuclear matter with fixed proton fractions and stellar
matter subject to $\beta$-equilibrium and trapped neutrinos. 
In homogeneous matter, $\alpha$ particles are only present at densities below 0.02 fm$^{-3}$ and their presence decreases with the increase of the 
temperature and, for a fixed temperature, the $\alpha$ particle fraction 
decreases for smaller proton fractions. A repulsive interaction is important 
to mimic the dissolution of the clusters in homogeneous matter. The effects 
of the $\alpha$ particles on the pasta structure is
very small except close to the critical temperatures and / or proton fractions 
when it may still predict a pasta phase while no pasta phase would occur in the
absence of light clusters. It is shown that for densities above 0.01 fm$^{−3}$ the $\alpha$ particle fraction in the pasta phase is much larger than the $\alpha$ particle fraction in homogeneous matter.
\end{abstract}
\maketitle

\vspace{0.50cm}
PACS number(s): {21.65.+f, 24.10.Jv, 26.60.+c, 95.30.Tg}
\vspace{0.50cm}

\section{Introduction}

The pasta phase is a frustrated system \cite{pethick,hashimoto,horo,watanabe,
maruyama} present
at densities of the order of 0.006 - 0.1 fm$^{-3}$ \cite{pasta1}
in neutral nuclear matter and 0.029 - 0.065 fm$^{-3}$ \cite{bao,pasta2} 
in $\beta$-equilibrium stellar matter, where a competition between the 
strong and the electromagnetic interactions takes place. 

The basic shapes of these complex structures were named \cite{pethick} 
after well known types of cheese and pasta: 
droplets = meat balls (bubbles = Swiss cheese), rods = spaghetti 
(tubes = penne) and slabs = lasagna, for three, two and one dimensions 
respectively. 
The pasta phase is the ground state configuration if its free energy per 
particle is lower than the corresponding to the homogeneous phase at the same 
density. 

A complete equation of state capable of describing matter ranging from very 
low densities to few times saturation density and from zero temperature to 
around 100 MeV is a necessary step towards the understanding of the stellar core
collapse, supernova explosion and protoneutron star evolution. The constitution
of the pulsar crust plays a definite role in the emission of neutrino and 
gravitational waves. In the inner crust of neutron stars (zero temperature, 
very low 
proton fraction, matter in $\beta$- equilibrium) the pasta phase is expected 
to be present and to coexist with a neutron gas. In supernova (finite 
temperature, proton fraction around 0.3) the pasta phase is structured in such 
a way that there is no neutron gas or it is very low in density 
\cite{watanabe05}. 

In previous works \cite{pasta1,pasta2} we have studied the 
existence of the pasta phase at zero and finite temperature within 
different parametrizations of the relativistic non-linear Walecka model 
(NLWM) \cite{sw}, namely NL3 \cite{nl3}, TM1 \cite{tm1} and GM3 \cite{glen} 
and  the onset of the pasta phase with different parametrizations of the 
density dependent hadronic model, namely TW \cite{tw}, DDH$\delta$ 
\cite{gaitanos} and GDFM \cite{gogelein,dalen} respectively. In both works
two different methods were used: the coexisting phases 
(CP), both at zero and finite temperature and the Thomas-Fermi (TF) 
approximation at zero temperature only. { It was found that matter in 
$\beta$-equilibrium presents a small (at zero temperature or small 
temperatures) or non-existent pasta structure (at finite temperature 
above $\sim$ 4 MeV) \cite{pasta1}. In fact, even for Skyrme forces it was 
shown that depending on the model, above 2-4 MeV there is no pasta phase for 
$\beta$-equilibrium matter \cite{camille08}.
If $\beta$-equilibrium is imposed,  the pasta phase 
{\bf could not be found} in a CP
calculation for the same surface energy parametrization used for fixed proton
fractions. This indicates the necessity to use a good parametrization for the
surface energy which is temperature, proton fraction and geometry dependent, 
as also stressed in \cite{gogelein,dalen}. The specific problem of an 
appropriate parametrization for the surface energy is tackled in the present
work.

In the above calculations we have assumed predefined shapes for the pasta clusters and  considered that the groundstate configuration is the one that minimizes the free energy.
In \cite{watanabe09} the authors could show by using ab-initio numerical simulations that in fact pasta phases can be formed in collapsing supernova. By compressing a bcc lattice of spherical nuclei it was shown that 
  an ordered structure of rod-like nuclei could be formed.

The importance of the $\alpha$ particles has been pointed out in the recent 
literature \cite{ls91,shen,roepke,hor06,blaschke09}. It is the most strongly bound system among all light systems
and it certainly plays a role in nuclear matter.
Lattimer and Swesty worked out the equation of state  (EOS) in the compressible extended liquid drop
model based on a non-relativistic framework \cite{ls91} appropriate for supernova
simulations, for a wide range of densities, proton fractions and temperatures, including the
contribution of $\alpha$-particle clusters.
 An excluded volume 
prescription is used to model the dissolution of $\alpha$ particles at high 
densities. 
The same is done by Shen {\em et al.} in \cite{shen,shenguides} where non-uniform matter composed of protons, 
neutrons, $\alpha$ particles and single species of heavy nuclei is described 
with the Thomas-Fermi approximation and the TM1 parametrization of the non-linear Walecka model
(NLWM). At low densities, these particles are
approximated by classical gases.
In \cite{nse} nuclear statistical equilibrium equations are 
calculated and the $\alpha$ particles are also taken into account.
In \cite{roepke} an effective four-body equation that includes self-energy 
corrections and Pauli blocking is done in a consistent way. The influence of 
cluster formation on nuclear matter EOS is calculated and the occurrence of
instabilities investigated in \cite{roepke03}. 
In \cite{hor06} the virial equation of state of low-density nuclear matter 
composed of neutrons, protons and $\alpha$-particles is presented and it is 
shown that 
the predicted $\alpha$ particle concentrations differ from the predictions of 
the EOS proposed in \cite{ls91} and \cite{shenguides}. 
The inclusion of small clusters in the EOS  is revisited
in \cite{blaschke09}, where the most important thermodynamical quantities are
calculated with light clusters up to the $\alpha$ particle with a density 
dependent relativistic model. The conditions for the liquid-gas phase transition
are obtained and it is seen how the binodal section is affected by the inclusion
of these clusters. Moreover, an EOS is obtained from the gaseous low densities 
with a clusterised matter to higher cluster-free homogeneous matter. 

In the present paper we investigate the influence of the $\alpha$ particles 
both on the homogeneous matter and on the onset, { size of the clusters}
and structure of the 
pasta phase within the relativistic mean field approximation. As all results
so far show model dependence, three different parametrizations are employed. 
We have chosen the NL3 \cite{nl3}, GM1 \cite{glen} parametrizations of the NLWM and the  TW
parametrization of the density dependent hadronic model \cite{tw}.
The calculations are carried out for fixed proton fractions and also for 
matter in $\beta$-equilibrium with trapped neutrinos. Although in previous
works it was found that matter in  $\beta$-equilibrium presents a small (at zero temperature) 
or non-existent pasta structure (at finite temperature) that is very sensitive to  the surface energy 
\cite{pasta1,pasta2}, once trapped neutrinos are present, the picture changes
considerably due to the large fraction of protons. According to studies on binodals and spinodals underlying the 
conditions for phase coexistence and phase transitions \cite{camille08,cphelena},
non-homogeneous matter with trapped neutrinos is expected to be found until 
temperatures around $T=12$ MeV, depending on the model considered. For this 
reason, we investigate this possibility next. 

The paper is organized as follows: in section II we briefly review the 
formalism underlying the homogeneous neutral $npe$ matter with the
inclusion of the $\alpha$ particles. 
In section III the pasta phase is built with the help of the coexisting phases
method and the prescription for the introduction of the $\alpha$s is given.
A complete study for the parametrization of the surface energy based on 
the Thomas-Fermi calculation is performed and presented in section IV. In section  V our results are 
displayed and commented and in section VI our conclusions are drawn.

\section{The Formalism}

We consider a system of protons and neutrons with mass $M$
interacting with and through an isoscalar-scalar field $\phi$ with mass
$m_s$,  a isoscalar-vector field $V^{\mu}$ with mass
$m_v$, an isovector-vector field  $\mathbf b^{\mu}$ with mass
$m_\rho$.  We also include $\alpha$ particles as bosons with mass $M_{\alpha}$
and a system of electrons with mass $m_e$. We do not consider models with 
$\delta$ mesons next, but their introduction is straightforward.
The Lagrangian density reads:

\begin{equation}
\mathcal{L}=\sum_{i=p,n}\mathcal{L}_{i}+\mathcal{L}_e\mathcal{\,+L}_{{\sigma }}%
\mathcal{+L}_{{\omega }}\mathcal{+L}_{{\rho }}\mathcal{+L}_{{\alpha }},
\label{lagdelta}
\end{equation}
where the nucleon Lagrangian reads
\begin{equation}
\mathcal{L}_{i}=\bar{\psi}_{i}\left[ \gamma _{\mu }iD^{\mu }-M^{*}\right]
\psi _{i}  \label{lagnucl},
\end{equation}
with 
\begin{eqnarray}
iD^{\mu } &=&i\partial ^{\mu }-\Gamma_{v}V^{\mu }-\frac{\Gamma_{\rho }}{2}{\boldsymbol{\tau}}%
\cdot \mathbf{b}^{\mu } - e \frac{1+\tau_3}{2}A^{\mu}, \label{Dmu} \\
M^{*} &=&M-\Gamma_{s}\phi,
\label{Mstar}
\end{eqnarray}
the $\alpha$ particles are described as in \cite{blaschke09} by
\begin{equation}
\mathcal{L}_{\alpha }=\frac{1}{2} (i D^{\mu}_{\alpha} \phi_{\alpha})^*
(i D_{\mu \alpha} \phi_{\alpha})-\frac{1}{2}\phi_{\alpha}^* M_{\alpha}^2
\phi_{\alpha},
\end{equation}
with
\begin{equation}
iD^{\mu } = i \partial ^{\mu }-\Gamma_{\alpha} V^{\mu },
\end{equation}
where
\begin{equation}
M_{\alpha}= 4 M - B_{\alpha}, \quad B_{\alpha}=28.3 ~~{\rm MeV},
\end{equation}
and $\Gamma_{\alpha}$ is included for mimicking
the excluded volume effect and consequent $\alpha$ particles dissolution at
high densities. In \cite{blaschke09} a more complete effective interaction for 
the  $\alpha$ particles is considered, including the $\sigma$ meson-$\alpha$ 
particle interaction and the Pauli shifts for the binding energy. In our 
approach we calculate a lower bound for the $\alpha$ particles present in the 
medium. As a reference we also consider the
opposite limit and take the $\alpha$-particles as a gas of free particles.
In most of our calculations $\Gamma_{\alpha}$ is either set to zero or to 
 $\Gamma_{\alpha}= 4 \Gamma_v$. We have also made tests using a lower value 
for this interaction ($2 \Gamma_v$) and the results are commented when 
appropriate.

The electron Lagrangian density is given by
\begin{equation}
\mathcal{L}_e=\bar \psi_e\left[\gamma_\mu\left(i\partial^{\mu} + e A^{\mu}\right)
-m_e\right]\psi_e,
\label{lage}
\end{equation}
and the meson Lagrangian densities are 
\begin{eqnarray*}
\mathcal{L}_{{\sigma }} &=&+\frac{1}{2}\left( \partial _{\mu }\phi \partial %
^{\mu }\phi -m_{s}^{2}\phi ^{2}-\frac{1}{3}\kappa \phi ^{3}-\frac{1}{12}%
\lambda \phi ^{4}\right)  \\
\mathcal{L}_{{\omega }} &=&\frac{1}{2} \left(-\frac{1}{2} \Omega _{\mu \nu }
\Omega ^{\mu \nu }+ m_{v}^{2}V_{\mu }V^{\mu }
+\frac{1}{12}\xi g_{v}^{4}(V_{\mu}V^{\mu })^{2} \right) \\
\mathcal{L}_{{\rho }} &=&\frac{1}{2} \left(-\frac{1}{2}
\mathbf{B}_{\mu \nu }\cdot \mathbf{B}^{\mu
\nu }+ m_{\rho }^{2}\mathbf{b}_{\mu }\cdot \mathbf{b}^{\mu } \right)
\end{eqnarray*}
where $\Omega _{\mu \nu }=\partial _{\mu }V_{\nu }-\partial _{\nu }V_{\mu }$
, and $\mathbf{B}_{\mu \nu }
=\partial _{\mu }\mathbf{b}_{\nu }-\partial _{\nu }\mathbf{b}
_{\mu }-\Gamma_{\rho }(\mathbf{b}_{\mu }\times \mathbf{b}_{\nu })$.
The  parameters of the models  are: the nucleon mass $M=939$ MeV,
the coupling parameters $\Gamma_s$, $\Gamma_v$, $\Gamma_{\rho}$ of the mesons to
the nucleons,  the electron mass $m_e$ and the electromagnetic coupling constant
$e=\sqrt{4 \pi/137}$.
In the above Lagrangian density $\boldsymbol {\tau}$ is
the isospin operator. When density dependent models are used, the non-linear
terms are not present and hence $\kappa=\lambda=\xi=0$ and the density dependent
parameters are chosen as in \cite{tw,gaitanos,inst04}. When models with
constant couplings are used, $\Gamma_i$ is replaced by $g_i$, where 
$i=s,v,\rho$ as in \cite{nl3,glen}. The bulk nuclear matter properties of the
models we use in the present paper are given in Table \ref{tab1}. We also 
include in the
table some properties at the thermodynamical spinodal surface: $\rho_{spin}$ 
is the upper border
density at the spinodal surface  for symmetric matter (it defines the density 
for which the
incompressibility is zero), $\rho_t$ and $P_t$ are, respectively, the density and the pressure at
the crossing between the cold $\beta$-equilibrium  equation of state and the spinodal
surface. They give a rough estimate of the density and pressure at the crust-core transition
\cite{pasta1,pasta2}. For symmetric matter or large proton fractions  we expect that GM1 will
have the largest extension of the pasta phase because it has the largest value of $\rho_{spin}$. For cold $\beta$-equilibrium  matter TW predicts
the largest pasta extension, with the largest transition density $\rho_t$.

\begin{table}[h]
\caption{ Nuclear matter properties at the saturation density and at the spinodal surface}
\label{tab1}
\begin{center}
\begin{tabular}{lccccccccc}
\hline
&  NL3 &  GM1 & TW \\
&   \cite{nl3} & \cite {glen} & \cite{tw} \\
\hline
\hline
$B/A$ (MeV) & 16.3  & 16.3  & 16.3 \\
$\rho_0$ (fm$^{-3}$) & 0.148  & 0.153  & 0.153 \\
$K$ (MeV) & 272  &  300 & 240 \\
${\cal E}_{sym.}$ (MeV)  & 37.4  & 32.5 & 32.8 \\
$M^*/M$ & 0.60 & 0.70  & 0.55 \\
$L$ (MeV) & 118.3 & 93.8  & 55.3 \\
$K_{sym}$ (MeV) & 100.5 & 17.9 & -124.7 \\
$Q_0$ (MeV) & 203 & -216& -540\\
$K_{\tau}$ (MeV) &-698 &-478&-332\\
$\rho_{spin}$ (fm$^{-3}$) &0.096 &0.100&0.096\\
$\rho_t$ (fm$^{-3}$) &0.0646&0.0743& 0.0850\\
$P_t$ (MeV/fm$^3$)& 0.396 &0.382& 0.455\\
\hline
\end{tabular}
\end{center}
\end{table}
From de Euler-Lagrange formalism we obtain coupled differential 
equations for
the scalar, vector, isovector-scalar and
nucleon fields. In the static case there are no currents and the spatial
vector components are zero. In \cite{pasta1} a complete description of the 
mean-field and Thomas-Fermi approximations applied to different 
parametrizations of the NLWM are given and we do not repeat them here.
The equations of motion for the  fields are obtained and solved 
self-consistently and they can be read off \cite{pasta1,pasta2}.
The above mentioned equations of motion depend on the 
the  equilibrium  densities $\rho =$ $\rho _{p}+\rho _{n}$,
$\rho_3 =\rho _{p}-\rho_{n} $ and  $\rho_{s} =$ $\rho _{sp}+\rho_{sn} $ 
where the proton/neutron densities  are given by
\begin{equation}
\rho_i=\frac{1}{\pi^2} \int {p^2 dp}(f_{i+}-f_{i-}),\,\, i=p,n
\end{equation}
and the corresponding scalar density by
\begin{equation}
\rho _{s_{i}}=\frac{1}{\pi^2} \int {p^2 dp}
\frac{M^{*}}{\sqrt{p^{2}+{M^{*}}^{2}}}(f_{i+}+f_{i-}),
\label{rhoscalar}
\end{equation}
with the distribution functions given by
\begin{equation}
f_{i \pm}=\frac{1}{1+\exp[(\epsilon^{\ast}({\mathbf p}) \mp \nu_i)/T]}\;,
\label{distf}
\end{equation}
where
${\epsilon}^{\ast}=\sqrt{{\mathbf p}^2+{M^*}^2}$,
\begin{equation}
M^*=M - \Gamma_s ~ \phi_0,
\label{effm}
\end{equation}
and the effective chemical potentials are
\begin{equation}
\nu_i=\mu_i - \Gamma_v V_0 - \frac{\Gamma_{\rho}}{2}~  \tau_{3 i}~ b_0
-{\Sigma^{R}_{0}} ,
\end{equation}
$\tau_{3 i}= \pm 1$ is the isospin projection for the protons and neutrons
respectively. The density dependent models in the mean field approximation 
contain a rearrangement term ${\Sigma^{R}_{0~i}}$ \cite{gaitanos} given by:
\begin{equation*}
{\Sigma^{R}_{0}}=
\frac{\partial \Gamma_{v}}{\partial \rho} (\rho) V_0 +
\frac{\partial \Gamma_{\rho}}{\partial \rho} \rho_3 ~ \frac{b_0}{2} -
\frac{\partial \Gamma_s}{\partial \rho} \rho_{s} \phi_0.
\end{equation*}

In the description of the equations of state
of a system, the required quantities are the baryonic density,  energy
density,  pressure and free energy, explicitly written in \cite{pasta1,pasta2}.
We refer next only to some of the quantities.

\noindent The free energy density is defined as:
\begin{equation}
{\cal F}= {\cal E}-T {\cal S},
\end{equation}

\noindent with the entropy density :
\begin{equation}
{\cal S}= \frac{1}{T}({\cal E}+P-\mu_p \rho_p - \mu_n \rho_n).
 \end{equation}

In the present work the $\alpha$ particles are included as bosons and their
chemical potential is obtained from the proton and neutron chemical potentials
as in \cite{nse,blaschke09}: 
\begin{equation}
\mu_{\alpha}= 2 (\mu_p + \mu_n).
\label{mua}
\end{equation}

 The inclusion of the $\alpha$ particles also give rise to a rearrangement term
given by
\begin{equation}
{\Sigma^{R}_{0 \alpha}}=
4 \frac{\partial \Gamma_{\alpha}}{\partial \rho} \rho_\alpha V_0, 
\end{equation}
which is present in the effective chemical potential of the $\alpha$ particles:
\begin{equation}
\nu_\alpha=\mu_\alpha - \Gamma_{\alpha} V_0 - \Sigma^{R}_{0 \alpha}.
\end{equation}

The density of $\alpha$ particles is 
\begin{equation}
\rho_{\alpha}=\frac{1}{\pi^2} \int {p^2 dp}(f_{\alpha+}-f_{\alpha-}),
\label{rhoalpha}
\end{equation}
with the boson distribution function given by
\begin{equation}
f_{\alpha \pm}\,=\,\frac{1}{-1+\exp[(\epsilon_{\alpha} \mp \nu_{\alpha})/T]},
\end{equation}
where $\epsilon_\alpha=\sqrt{p^2+M_{\alpha}^2}$.
The $\alpha$ free energy density reads
\begin{equation}
{\cal F}_{\alpha}= {\cal E}_{\alpha}-T {\cal S}_{\alpha},
\end{equation}
where ${\cal E}_{\alpha}$ and ${\cal S}_{\alpha}$ stand respectively for 
bosonic energy and entropy density.  

It is worth emphasizing that a term
proportional to $\Sigma^{R}_{0  \alpha} \rho_\alpha$ is present in the pressure.
When non linear Walecka type models are used there is no rearrangement term and 
the $\alpha$ effective chemical potential is also given by eq.(\ref{mua}).
{For the temperatures we consider in the present work no alpha particle
condensation occurs, and, therefore, the above equations do not contain
the condensate contribution.}

As for the electrons, their density and distribution functions read:
\begin{equation}
\rho_e=\frac{1}{\pi^2} \int {p^2 dp}(f_{e+}-f_{e-}),
\label{rhoe}
\end{equation}
with
\begin{equation}
f_{e\pm}({\mathbf r},{\mathbf p},t)\,=\,\frac{1}{1+\exp[(\epsilon_e\mp
\mu_e)/T]},
\end{equation}
where $\mu_e$ is the electron chemical potential and
$\epsilon_e=\sqrt{p^2+m_e^2}$. 
We always consider neutral matter and therefore the electron density is
equal to the {charge  density of the positive} charged particles (protons and alphas) . 

The free energy density of the electrons reads
\begin{equation}
{\cal F}_e= {\cal E}_e-T {\cal S}_e,
\end{equation}
with
\begin{equation}
{\cal S}_e= \frac{1}{T}({\cal E}_e+P_e-\mu_e \rho_e ).
 \end{equation}

\section{Coexisting phases}

Two possibilities are discussed next: nuclear matter with fixed proton fraction
and stellar matter with $\beta$-equilibrium and trapped neutrinos.

\subsection{Nuclear pasta}

As in \cite{pasta1,maruyama}, for a given total density $\rho$ and proton 
fraction, now defined with the inclusion of the protons present inside the
$\alpha$ particles, the pasta structures are built with different 
geometrical forms in a background nucleon gas. This is achieved by calculating 
from the Gibbs' conditions the  
density and the proton fraction of the pasta and
of the background gas, so that in the whole we have to  
 solve simultaneously the following six 
equations:
\begin{equation}
P^I(\nu_p^I,\nu_n^I,{M^*}^I)=P^{II}(\nu_p^{II},\nu_n^{II},
{M^*}^{II}),\label{gibbs1}
\end{equation}
\begin{equation}
\mu_i^I=\mu_i^{II}, \quad i=p,n\label{gibbs2}
\end{equation}
\begin{equation}
m_s^2 \phi_0^I + \frac{\kappa}{2} {\phi_0^2}^I 
+\frac{\lambda}{6}{\phi_0^3}^I = g_s \rho_s^I,\label{gibbs3}
\end{equation}
\begin{equation}
m_s^2 \phi_0^{II} + \frac{\kappa}{2} {\phi_0^2}^{II} 
+\frac{\lambda}{6}{\phi_0^3}^{II} = g_s \rho_s^{II},\label{gibbs4}
\end{equation}

\begin{equation}
f (\rho_p^I + 2 \rho_{\alpha}^I) + (1-f) (\rho_p^{II} + 2 \rho_{\alpha}^{II})
= Y_p \rho,\label{gibbs7}
\end{equation}
where I and II label each of the phases, $f$ is the volume fraction of 
phase I: 
\begin{equation}
f= \frac{\rho -\rho^{II}}{\rho^I-\rho^{II}}, 
\end{equation}
where the total baryonic density is
\begin{equation}
\rho=\rho_p + \rho_n + 4 \rho_{\alpha},
\end{equation}
and $Y_p$ is the global proton fraction given by
\begin{equation}
Y_p = \frac{\rho_p + 2 \rho_{\alpha}}{\rho}.
\end{equation}

The density of electrons is uniform and taken as $\rho_e=Y_p \rho$. 
The total pressure is given  by $P=P^I+P_e+P_{\alpha}$. 
The total energy density of the system is given by
\begin{equation}
{\cal E}= f {\cal E}^I + (1-f) {\cal E}^{II} + {\cal E}_e +
{\cal E}_{surf} + {\cal E}_{Coul}, 
\label{totener}
\end{equation}
where, by minimizing the sum  ${\cal E}_{surf} + {\cal E}_{Coul}$ with respect
to the size of the droplet/bubble, rod/tube or slab we get 
\cite{maruyama}
${\cal E}_{surf} = 2 {\cal E}_{Coul},$ and  
\begin{equation}
{\cal E}_{Coul}=\frac{2 F}{4^{2/3}}(e^2 \pi \Phi)^{1/3} 
\left(\sigma D (\rho_p^I-\rho_p^{II})\right)^{2/3},
\end{equation}
where $F=f$ for droplets and $F=1-f$ for bubbles, 
 $\sigma$ is the surface energy coefficient,
$D$ is the dimension of the system. For droplets, rods and slabs,
\begin{equation}
\Phi=\begin{cases}
\left(\frac{2-D f^{1-2/D}}{D-2}+f \right) \frac{1}{D+2}, \quad D=1,3;\\
 \frac{f-1-ln(f)}{D+2}, \quad D=2. \end{cases}
\end{equation}
and for bubbles and tubes the above expressions are valid with 
$f$ replaced by $1-f$.

{
Each structure is considered to be in the center of a charge neutral 
Wigner-Seitz  cell
constituted by neutrons, protons and leptons \cite{shen}. 
The Wigner-Seitz cell
is a sphere/cilinder/slab whose volume is the same as the unit BCC cell. 
In \cite{shen} the
internal structures are associated with heavy nuclei. Hence, the radius of the
droplet (rod,slab) and of the Wigner-Seitz cell are
respectively given by:
\begin{equation}
R_D=\left( \frac{\sigma D}{4 \pi e^2 (\rho_p^I+2 \rho_{\alpha}^I-\rho_p^{II}
- 2 \rho_{\alpha}^{II})^2 \Phi} \right)^{1/3},
\end{equation}
\begin{equation}
R_W=\frac{R_D}{(1-f)^{1/D}}. 
\end{equation}
  
In Fig. \ref{sig} the surface energy is plotted as a function of the proton 
fraction for the three models under study. 
It is seen that the models differ a lot.}

\subsection{Surface energy}

{ The authors of \cite{watanabe2001} have studied how the uncertainties on the surface energy affect the appearance of non-spherical pasta 
structures. In particular, they have shown that for typical values of the surface energy
non-spherical clusters may occur below the transition density to uniform matter.} Also the authors of \cite{maruyama} state that the appearance of 
the pasta phase essentially depends on the value of the surface tension. We 
have fixed the surface tension at different values
and confirmed their claim in \cite{pasta1,pasta2}, where the surface energy 
coefficient was parametrized in terms of the proton fraction
according to the functional proposed in  \cite{lattimer}, obtained
by fitting Thomas-Fermi and Hartree-Fock numerical values with a Skyrme force.

\begin{figure}
\begin{center}
\includegraphics[width=0.9\linewidth,angle=0]{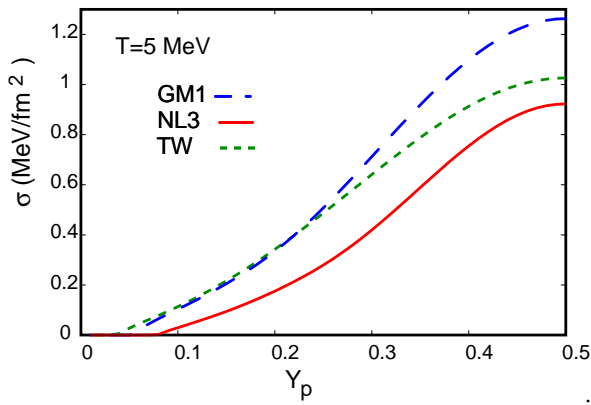} \\
\end{center}
\caption{(Color online) Surface energy as a function of the proton fraction for GM1, NL3 and TW-}
\label{sig}
\end{figure}

In the present paper we have considered this point in a more systematic and
consistent way. Hence, we  have parametrized the surface energy according to
our Thomas-Fermi calculations for the relativistic models under investigation.
We have used the Gibbs prescription to obtain the $\sigma$ coefficient 
which, as extensively discussed in the literature \cite{centel,mayers}, is the
appropriate surface tension coefficient to be used. We have obtained for the 
three parametrizations used in this work, namely NL3, TW and GM1 
parameter sets, the corresponding fittings for $\sigma$. Here, we briefly
discuss the method that we have employed for fitting $\sigma$ and more specific
details about the numerical algorithm can be found in \cite{pasta1}. 
First, we obtain a density profile for the two component system consisting 
of protons and neutrons in a 
semi-infinite one-dimensional system, where the axis perpendicular to the
interface has been  defined as the z-axis. 
In our approach, we assume the system inside a large box of
radius R so that the densities at the left and right boundaries correspond to 
asymptotic densities that approach the values of uniform nucleon 
matter (nucleus) in equilibrium with a gas of drip particles and
a uniform gas of drip  particles respectively. 
  Hence, we need a recipe in order to extract the surface energy from the bulk
  one. The surface tension may be expressed as follows:
\begin{equation}
\sigma =\int_{-\infty}^{\infty} dz \left[\epsilon (z) - \epsilon_d -
\epsilon_{ref} (z)   \right]
\end{equation}
as discussed in \cite{centel}. The quantity $\epsilon_{ref} (z)$ corresponds to a reference
energy density associated to the bulk contribution, $\epsilon_d =
\lim_{z \rightarrow \infty}\epsilon(z)$ and $\epsilon(z)$ is the energy
density. If one considers the Gibbs phase coexistence conditions and
usual thermodynamic relations  one obtains for the (Gibbs) surface tension
coefficient the expression:  
$$\sigma =\int_{-\infty}^{\infty} dz \left( \epsilon (z) - \epsilon_d - 
\mu_n \left[ \rho_n(z)-\rho_{nd} \right] \right.$$
$$\left. - \mu_p \left[ \rho_p(z)-\rho_{pd} \right]  \right) ~~,$$
where $\rho_{nd}=\lim_{z\rightarrow \infty} \rho_n(z)$ and an analogous
definition for $\rho_{pd}$. 
 An alternative and completely equivalent expression for the calculation of
 $\sigma$, known as thin-wall approximation \cite{marina,cpsig,dmcp} is given
 by the  following expression:
\begin{equation}
\sigma = \int_{0}^{\infty} dz \left( \left(\frac{d\phi_0}{dr}\right)^2 
-\left(\frac{dV_0}{dr}\right)^2 
- \left(\frac{db_0}{dr}\right)^2 \right)  ~. \label{sigcp}
\end{equation}
In our parametrization of $\sigma$ the above expression has been used since 
it is more convenient for our Thomas-Fermi numerical
algorithm \cite{pasta1}. In particular,
it is specially adequate for the parametrization of the dependence of $\sigma$ 
on the temperature.
We do not try to 
parametrize the $\sigma$ through standard dimensionless
functions which, as discussed in \cite{mayers}, do not give good fits for
relativistic mean field models, specially, for the region of small
$\sigma$ values which are important for the pasta study. So, we follow a more 
pragmatic way  adopting a mathematical formula for $\sigma$ that gives accurate 
results for a broad  range of neutron excess and for temperatures up to 10MeV, 
which are adequate for the studies addressed in the present work. The
following functional for the surface tension coefficient, $\sigma$, is used,  
\begin{equation}
\sigma = \sigma (x,T=0)\left[ 1-a(T)~x \rm{T} -b(T)T^2 \right]~, \label{sigpar}
\end{equation}
where  $x=\delta^2$  stands for the squared relative neutron excess:
$$\delta = \frac{\rho_n-\rho_p}{\rho} = 1-2 Y_p~,$$
{ where  $\rho$, $\rho_n$, and $\rho_p$ are defined at $z=-\infty$.}
A table with the coefficients $\sigma (x,T=0)$, $a(T)$ and $b(T)$ for the NL3, 
TW and GM1 parametrizations is shown in the appendix.
The proton fraction considered throughout the calculation of $\sigma$ is the 
one of the denser phase.

\subsection{Stellar pasta for matter in $\beta$ equilibrium with trapped 
neutrinos}

In this case, hadronic matter is in $\beta$ equilibrium and the electron
neutrinos are trapped.
The condition of $\beta$ equilibrium in a system of protons, neutrons,
electrons and neutrinos is 
\begin{equation}
\mu_p=\mu_n-\mu_e+\mu_{\nu_e}.
\end{equation}
Besides the Gibbs conditions given in eqs. (\ref{gibbs1}), (\ref{gibbs2}),
(\ref{gibbs3}) and (\ref{gibbs4}) we also impose
\begin{equation}
Y_l = \frac{\rho_e+\rho_{\nu_e}}{\rho},
\end{equation}
where $Y_l$ is the lepton fraction and has been fixed as 0.4 and 
$\rho_{\nu_e}$ is the density of electron neutrinos given by
\begin{equation}
\rho_{\nu_e}=\frac{1}{2 \pi^2} \int {p^2 dp}(f_{\nu_{e+}}-f_{\nu_{e-}}),
\label{rhonue}
\end{equation}
with
\begin{equation}
f_{\nu_{e \pm}}=\,\frac{1}{1+\exp[(p \mp \mu_{\nu_e})/T]}.
\end{equation}
The neutrino free energy density reads
\begin{equation}
{\cal F}_{\nu_e}= {\cal E}_{\nu_e}-T {\cal S}_{\nu_e},
\end{equation}
where ${\cal E}_{\nu_e}$ and ${\cal S}_{\nu_e}$ stand respectively for 
the neutrino energy and entropy density.  

The densities of interest to the study of the pasta phase are too low for 
the muons to appear, which generally occur for densities above 
$0.1$ fm$^{-3}$ \cite{camille08} and hence are not considered in the present 
work.

\section{Results and discussions}

In this section we show results for the fraction of $\alpha$-particles in 
homogeneous
matter and pasta-like matter and discuss the effect of the  
$\alpha$-particles on the
structure of the pasta phase.

\subsection{$\alpha$ particles in homogeneous matter}
\begin{figure}
\begin{center}
\includegraphics[width=0.9\linewidth,angle=0]{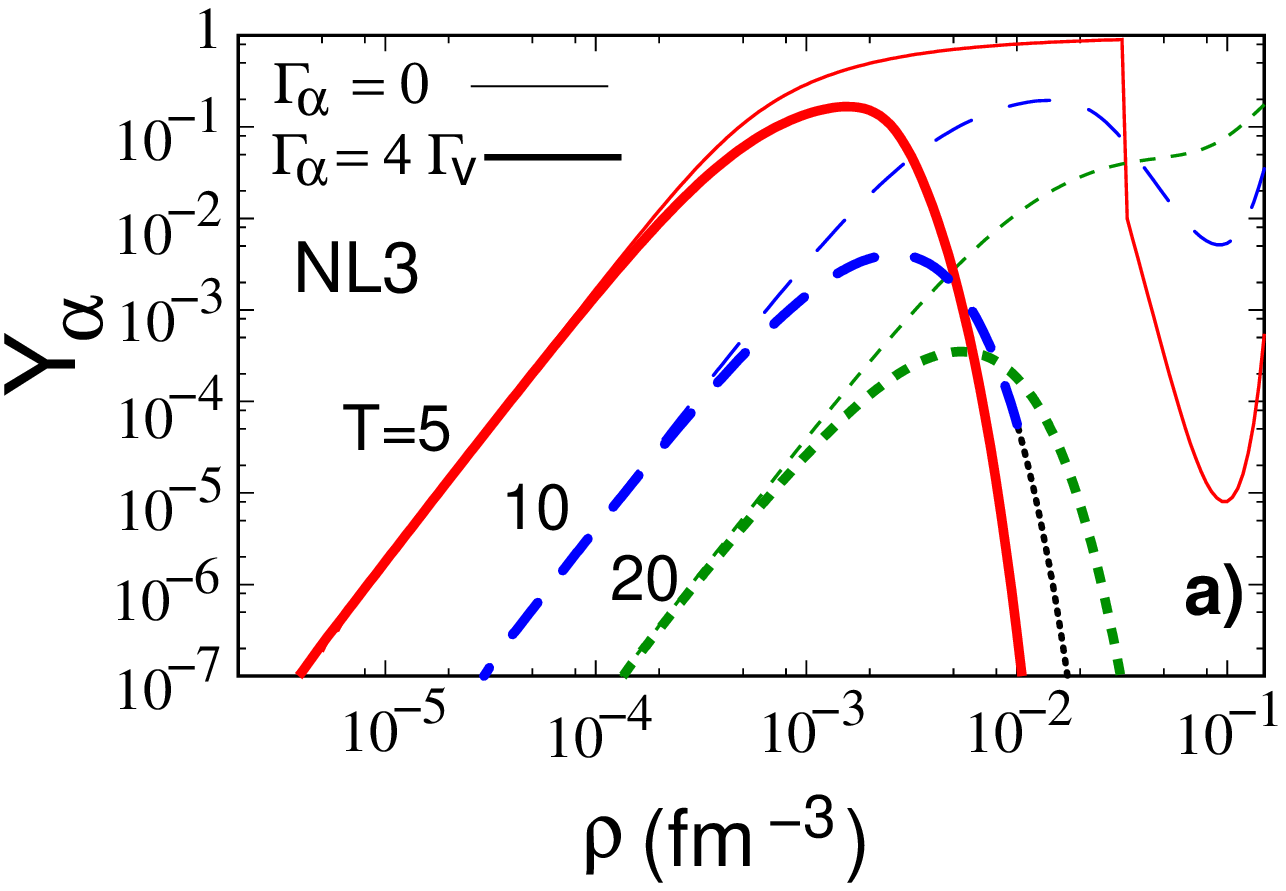} \\
\includegraphics[width=0.9\linewidth,angle=0]{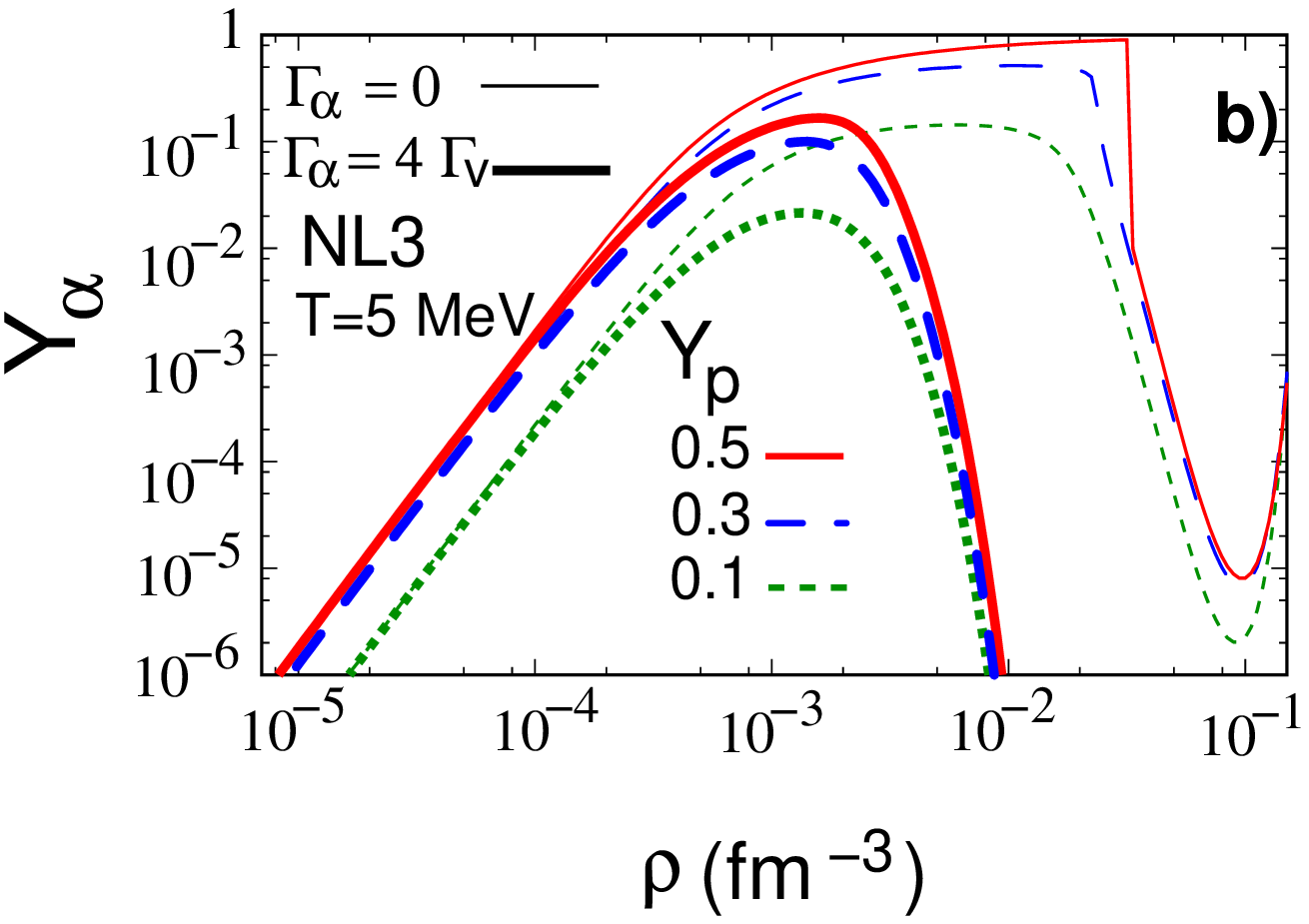} 
\end{center}
\caption{(Color online) $\alpha$ fractions  as a function of density for  a)
 symmetric matter and several  temperatures and  b)  T=5 MeV and
 several proton
fractions, obtained with NL3. The thin curves are for free $\alpha$-particles and the thick ones
include the $\omega$-meson-$\alpha$ particle interaction.}
\label{fig1}
\end{figure}

We show the amount of $\alpha$ particles present
in homogeneous nuclear matter described by the NL3 parametrization for 
symmetric matter and different temperatures in Fig.\ref{fig1}a 
and for $T=5$ MeV and three proton
fractions, $Y_p=0.5,\,0.3,\,0.1$ in Fig. \ref{fig1}b. 
In all figures 
the curves drawn with thin lines were obtained
for free $\alpha$-particles, $\Gamma_{\alpha}=0$ 
and the curves with thick lines for $\Gamma_{\alpha}= 4 \Gamma_v$, i.e., with the inclusion of the term that 
mimics the excluded volume  factor. As mentioned before, this term
dissolves the clusters. In both calculations the $\alpha$-particle vacuum 
mass was used. The maximum of the distribution occurs for the
density that maximizes  $2(\mu_p+\mu_n)-\Gamma_\alpha V_0-M_\alpha$.
The curves obtained with this term show 
the same behavior found in Fig. 15 of \cite{blaschke09}, i.e., the $\alpha$ 
particle fraction decreases with the increase of the temperature.
One can see from Fig.\ref{fig1}b that the total proton fraction $Y_p$ has a 
different effect on the $\alpha$ particle distribution. 
For $\Gamma_{\alpha}= 4 \Gamma_v$,
the maximum of the distribution occurs at the same density for all fractions,
although the amount of $\alpha$ particles shows a lower value for the 
smaller value of $Y_p$. Moreover, 
the density of dissolution of 
$\alpha$ particles does not seem to be sensitive to the proton fraction, 
because it is defined by the isoscalar vector interaction.
{\bf 
The $\alpha$ particle frations for free $\alpha$ particles in
  Fig. \ref{fig1}a and \ref{fig1}b 
show discontinuities  for large proton fractions at T=5 MeV, just above $\rho=0.02$
fm$^{-3}$. For lower proton fractions no discontinuity occurs.  
The region where the discontinuity occurs is a region of instability since the free energy has
a negative concavity. For T=5 MeV  and  $y_p$=0.3 there is a smooth transition from a region with a large alpha
particle fraction to a region with low alpha particle fraction.  If we increase the proton
fraction the transition is not any more smooth and we obtain a jump in the $\alpha$ particle
fraction.
The discontinuity occurs when the alpha particle 
fraction becomes close to one  and the chemical potential of the $\alpha$-particle
  close to the  alpha
particle mass. 
For
larger densities the nucleon chemical potential decreases due to the
attraction induced by the $\sigma$ meson. The minimum of the nucleon
chemical potential at $\rho\sim 0.1$ fm$^{-3}$  corresponds to the
minimum of the alpha particle fraction. The discontinuity or kink
occurs at the  density, or slightly above,  for which the
nucleon  chemical potential in the presence of alpha particles becomes
equal to the nucleon  chemical potential without alpha particles.
  To correctly treat this region of densities, where the free energy has a negative concavity,  we should consider that matter is not
homogeneous and there are two phases. However, describing alpha particles as free particles is
not realistic and when the interaction between particles is included the discontinuity in the
$\alpha$ particle fraction does not occur.}

\begin{figure}
\begin{center}
\includegraphics[width=0.9\linewidth,angle=0]{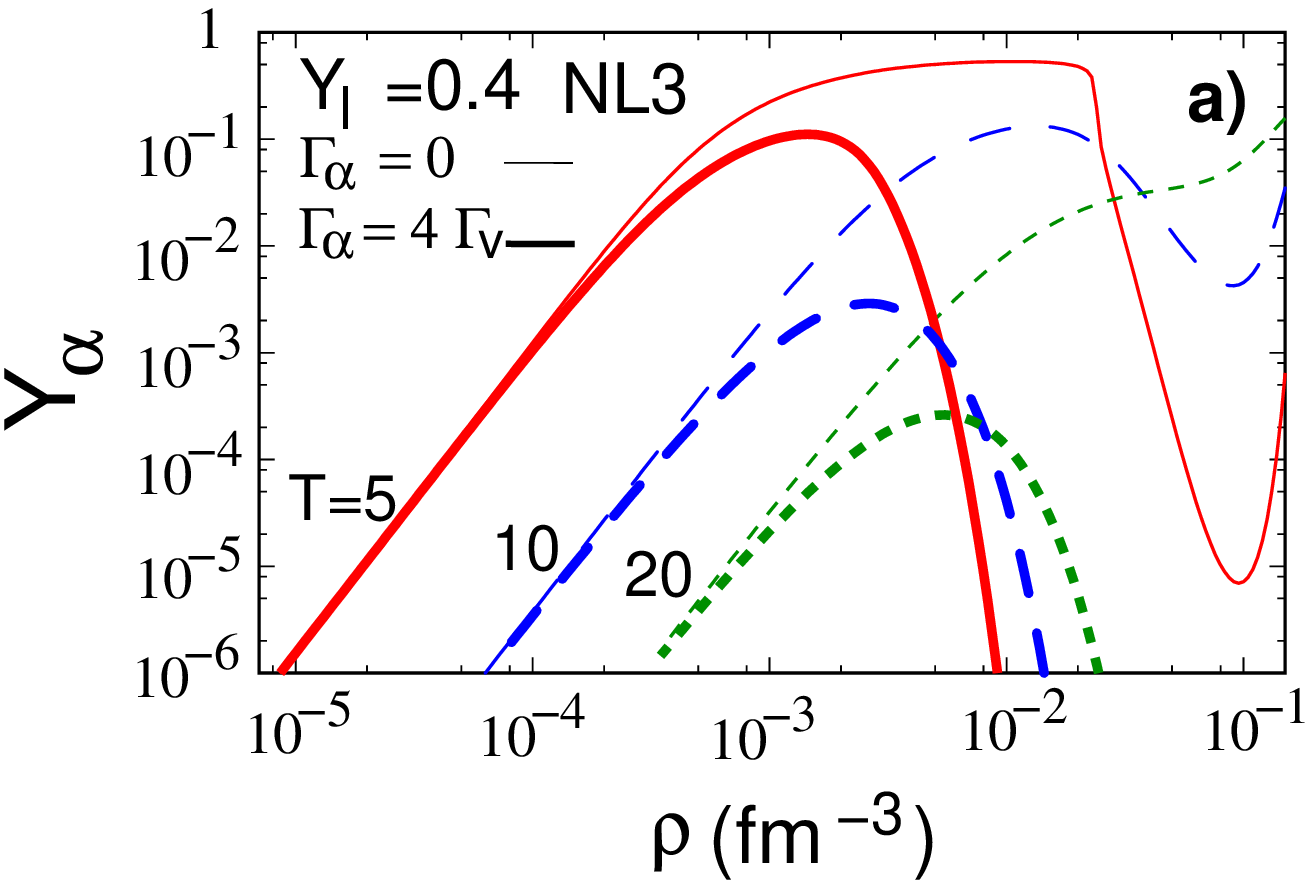} \\
\includegraphics[width=0.9\linewidth,angle=0]{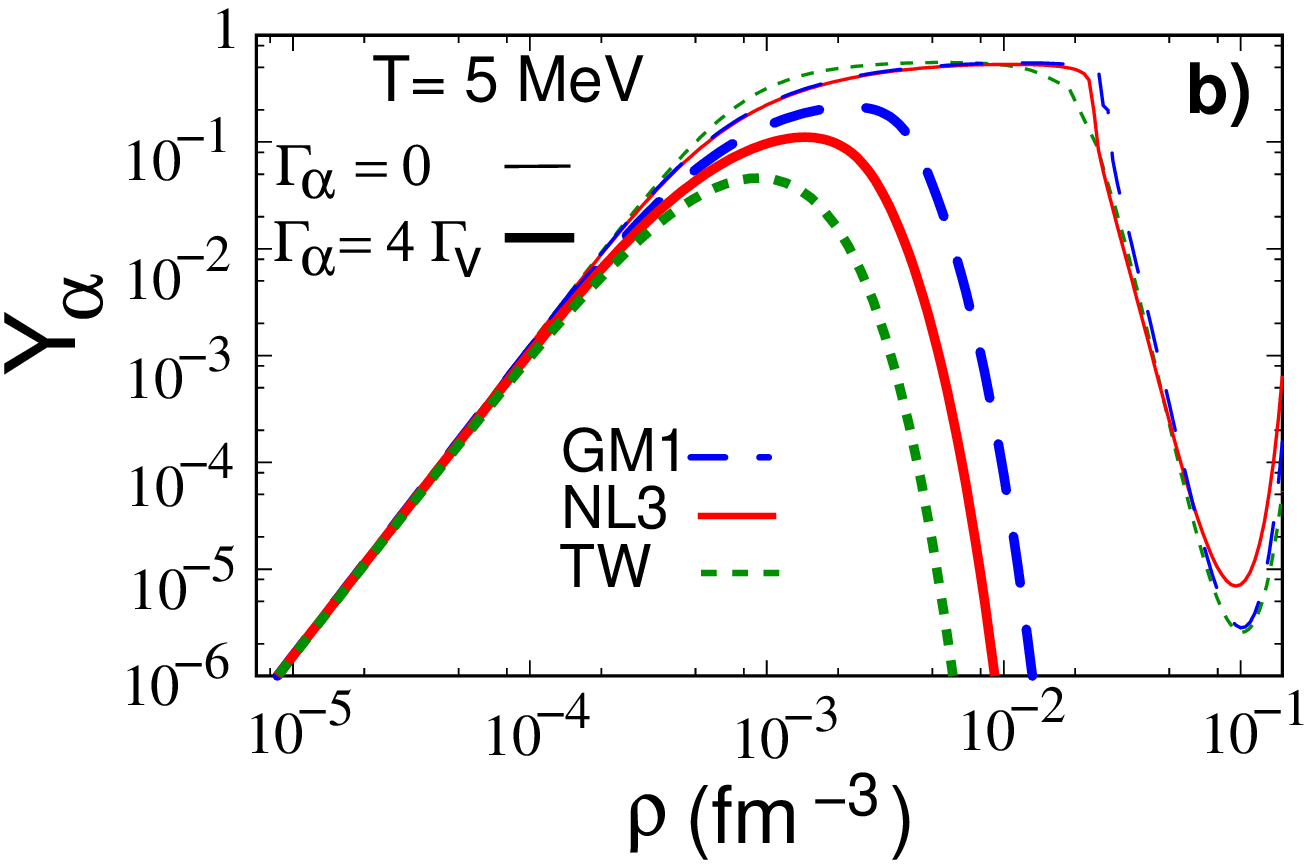} \\
\includegraphics[width=0.9\linewidth,angle=0]{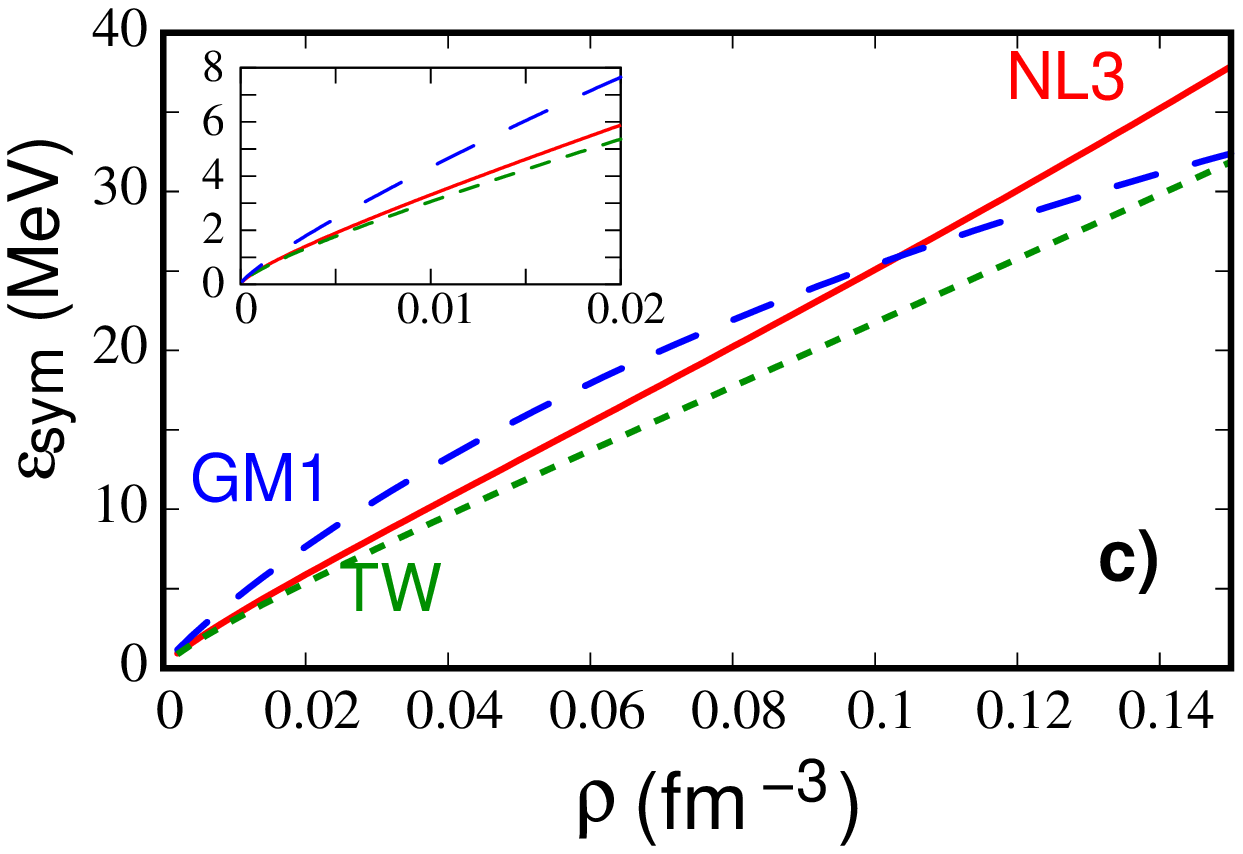} \\
\end{center}
\caption{(Color online) $\alpha$ fractions as a function of density obtained with a)  
model NL3 for different temperatures; b) models
 NL3, GM1 and TW for $T=5$ MeV for matter in $\beta$-equilibrium with 
trapped neutrinos. The thin curves are for free $\alpha$-particles and the 
thick ones
include the $\omega$-meson-$\alpha$ particle interaction.
In c) the { nucleon effective mass} within NL3, GM1 and TW is plotted at 
subsaturation densities.}
\label{fig2}
\end{figure}

In Fig. \ref{fig2}a the amount of $\alpha$ particles present in matter with
$\beta$-equilibrium and trapped neutrinos with the NL3 parametrization 
is displayed for different temperatures. Just like in  Fig. \ref{fig1}a),
it is seen that  the larger the temperature the lower the maximum of the 
$\alpha$ particle distribution. With the increase of temperature, the maximum 
is shifted to larger densities as well as the dissolution of the clusters. 
This behavior was also obtained in \cite{blaschke09}. However in  
\cite{blaschke09} other smaller clusters, besides the $\alpha$ particles, 
have also been considered. 
In Fig. \ref{fig2}b) the same is shown for T=5 MeV and the three models
used in the present work. 
TW is responsible for the smallest amount of $\alpha$ particles, whilst GM1 
for the largest. These behaviors are defined by the model
properties at subsaturation densities. This can be seen in Fig. \ref{fig2}c) 
{ where the nucleon effective mass of NL3, GM1 and TW are plotted for
subsaturation densities.  If the nucleon mass is larger it becomes
energetically more favorable to produce more alpha particles and less free 
nucleons. Both the
cluster dissolution and the α particle maximum occur at
smaller densities for TW and larger for GM1 for the same
reason.}


The virial expansion gives a model 
independent prediction of  the low density limit of the EOS 
\cite{hor06}. In Fig. \ref{virial} we compare the $\alpha$ particle fraction 
obtained with  the
virial EOS of low density nuclear matter \cite{hor06} with the prediction of 
the models GM1, NL3 and TW with interacting $\alpha$ particles, $\Gamma_\alpha=4 \Gamma_v$. We
also include the NL3 results for free $\alpha$ particles (thin red line). The GM1
parametrization with interacting $\alpha$ particles gives the closest  results to the virial
expansion and TW deviates the most.

\begin{figure}
\begin{center}
\includegraphics[width=0.9\linewidth,angle=0]{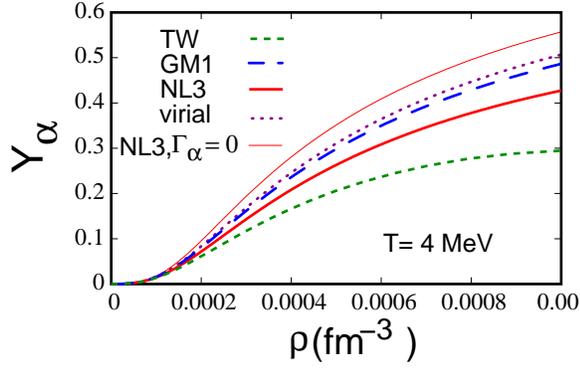} \\
\end{center}
\caption{(Color online) $\alpha$ fractions as a function of density obtained with GM1,  
NL3 and TW models are
compared with the virial EOS results of ref. \cite{hor06}.}
\label{virial}
\end{figure}

\subsection{$\alpha$ particles in the pasta phase}

\begin{figure}
\begin{center}
\includegraphics[width=0.9\linewidth,angle=0]{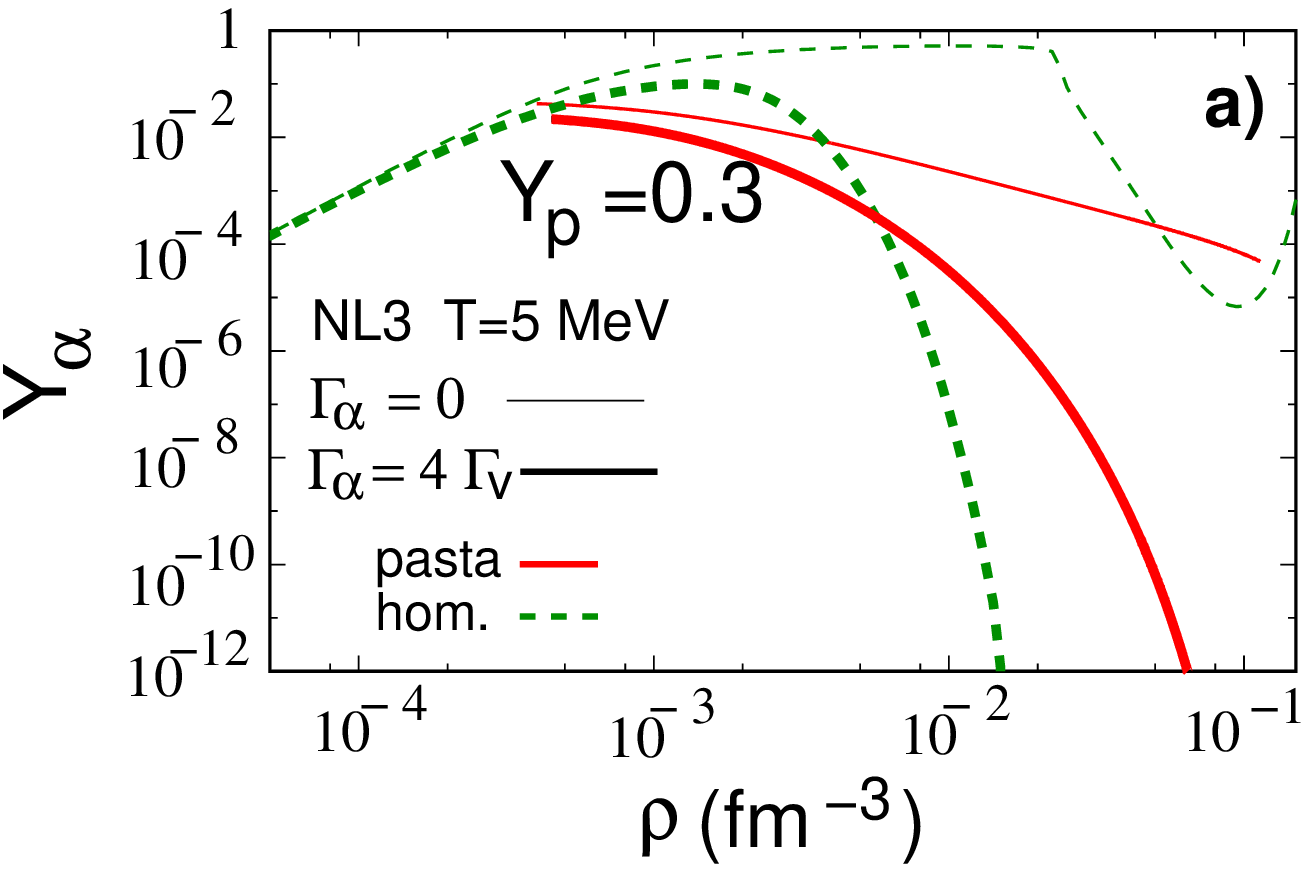} \\
\includegraphics[width=0.9\linewidth,angle=0]{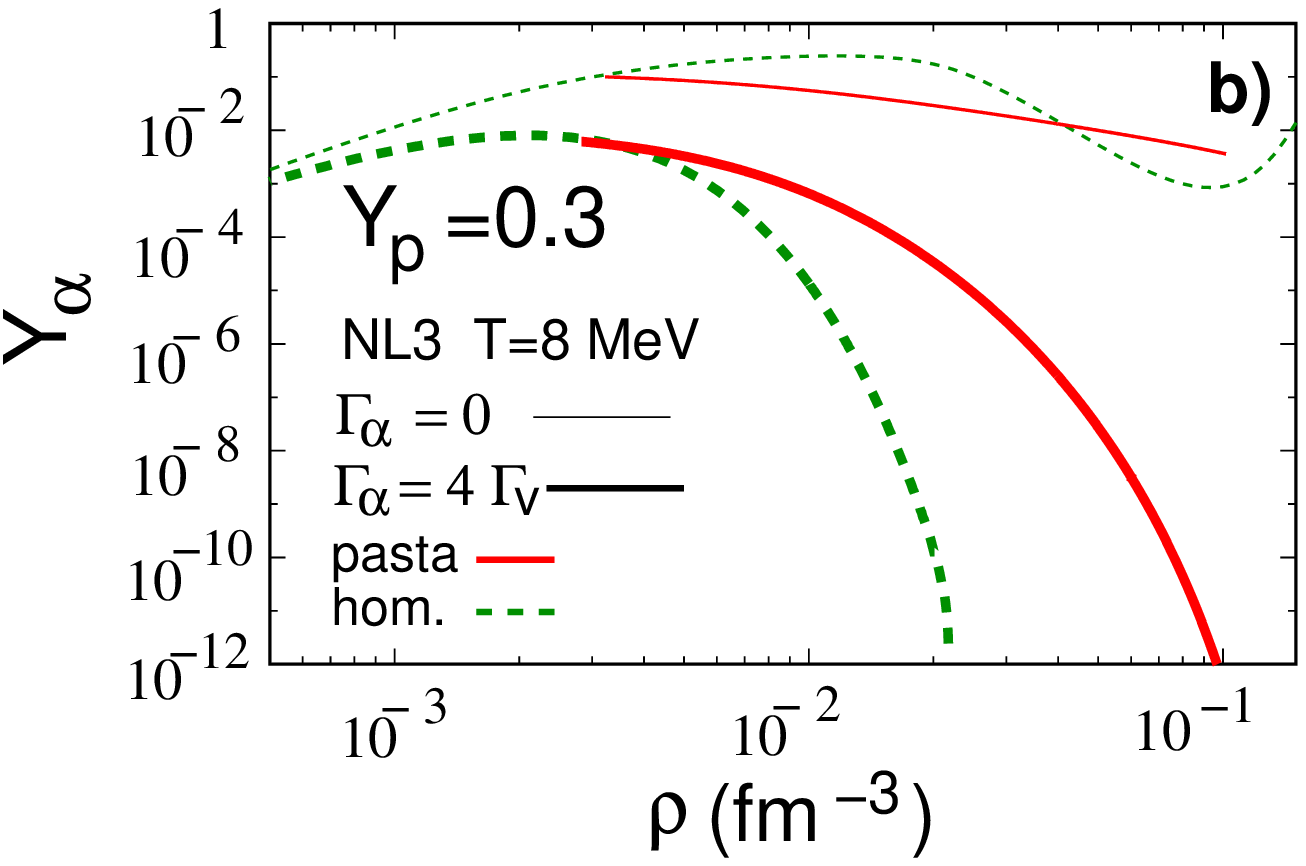} \\
\end{center}
\caption{(Color online) $\alpha$ fractions as a function of density obtained with  model NL3 for a) $Y_p=0.3$
and $T=5$ MeV; b)   for $Y_p=0.3$ and  $T=8$ MeV.
The thin curves are for free $\alpha$-particles and the thick ones
include the $\omega$-meson-$\alpha$ particle interaction.}
\label{fig3}
\end{figure}

\begin{figure}
\begin{center}
\includegraphics[width=0.9\linewidth,angle=0]{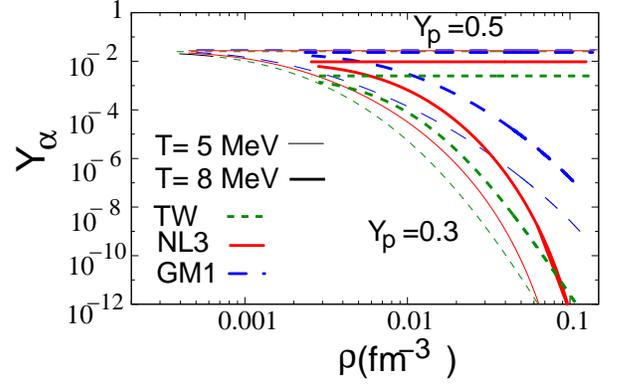} \\
\end{center}
\caption{(Color online) $\alpha$ fractions as a function of density obtained with  
{ the models under investigation} for $Y_p=0.5$ and
$Y_p=0.3$ and $T=5, \,8$ MeV.
The thin curves are for T=5 MeV and the thick curves for T=8 MeV}
\label{fig3c}
\end{figure}

In the sequel, we analyse the effect of the pasta phase on the fraction of
$\alpha$-particles. The $\alpha$ particles are present in the background gas which surrounds the pasta structures.
  \begin{figure}
\begin{center}
\includegraphics[width=0.9\linewidth,angle=0]{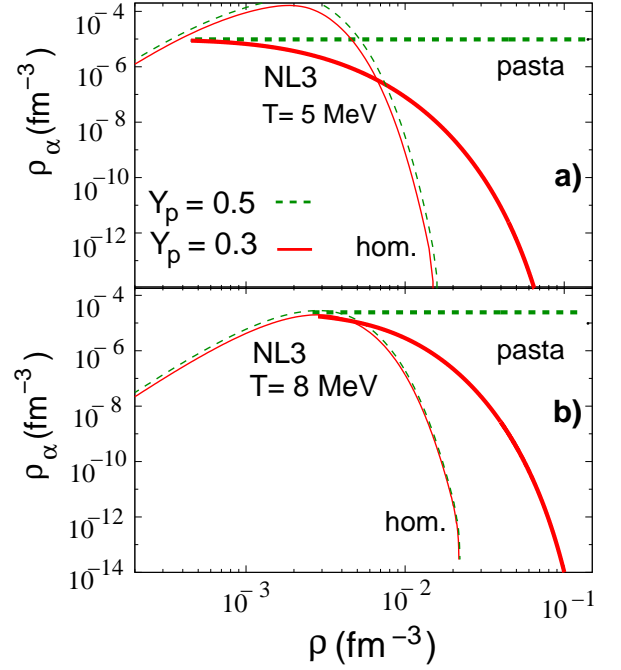} \\
\end{center}
\caption{(Color online) $\alpha$ particle density for a)  T=5 MeV and b) T= 8 MeV and  $Y_p=0.3$ and 0.5 obtained with  NL3 for homogeneous
  matter (thin lines) and gas phase of the pasta-like matter (thick lines). All
calculations include the $\omega$-meson-$\alpha$ particle interaction.}
\label{fig4}
\end{figure}

\begin{figure}
\begin{center}
\includegraphics[width=0.9\linewidth,angle=0]{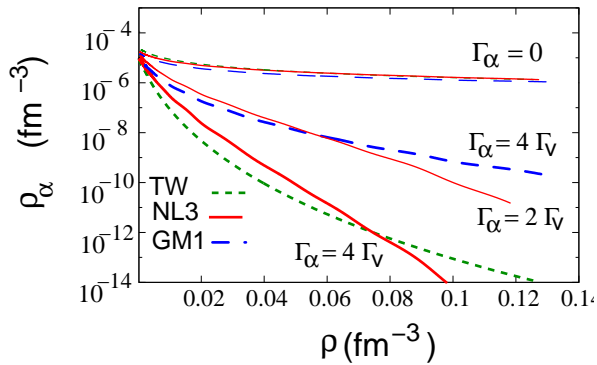} \\
\end{center}
\caption{(Color online) $\alpha$ particle density  
in the gas phase of pasta-like
matter for T=5 MeV  and $\beta$-equilibrium matter with trapped neutrinos
 obtained with  NL3, GM1 and TW. The $\alpha$ particle density is given for 
three cases: free
$\alpha$ particles $\Gamma_\alpha=0$ (thin lines),  interacting 
$\alpha$-particles with
$\Gamma_\alpha= 4\Gamma_v$ (thick lines), and just for NL3, 
interacting $\alpha$-particles with $\Gamma_\alpha= 2\Gamma_v$ (red medium thick 
line).}
\label{fig5}
\end{figure}

In Fig \ref{fig3} we have plotted the $\alpha$ particle fraction
$Y_\alpha$ for T=5 and  8 MeV with  the proton fraction $Y_p=0.3$. 
 Notice that $Y_\alpha=4 \rho_\alpha/\rho$ {  is the fraction of nucleons clustered in $\alpha$-particles.} In both 
figures we include the homogeneous matter
results (dashed lines) and the pasta-like matter results (full lines). We 
have considered both
free $\alpha$ particles (thin lines) and interacting $\alpha$ particles 
(thick lines). 
If free $\alpha$ particles are considered the fraction $Y_\alpha$  
in homogeneous matter becomes quite high and it may be close to one which is 
quite unrealistic. In pasta-like matter {the maximum of} $Y_\alpha$
is smaller, but still quite high: it varies between $\sim$ 0.01 and
0.1 and, for $Y_p=0.3$, it is larger for 
larger temperatures, $T=8$ MeV. This last tendency is still true when we include the interaction
of $\alpha$ particles with the $\omega$ meson. If we analyze the results including the
$\alpha-\omega$ interaction some important conclusions can be drawn: while for homogeneous
matter $Y_\alpha<10^{-12}$ for $\rho>0.02$ fm$^{-3}$, for pasta-like matter this only occurs
 at the crust-core boundary for   $\rho\sim 0.1$ fm$^{-3}$. 

We may consider the prediction of Fig. \ref{fig3} a  lower bound prediction 
since for
the $\alpha$ particle interaction we have only considered a repulsive interaction which mimics
the volume exclusion effect. If attraction would have also been considered we would expect
larger fractions \cite{blaschke09}. Also we have taken for the coupling constant the
nucleon-meson coupling constant multiplied by the mass number and this could be a too strong coupling.
In pasta-like matter for symmetric matter the $\alpha$ particle-fraction does not change
due to the method used to calculate the pasta phase. Phase equilibrium for symmetric nuclear
matter predicts the same proton fraction, equal to 0.5, as well as the same pressure for both
the dense (liquid) and gas phases of the pasta. We may expect that within a Thomas-Fermi
calculation the  $\alpha$-particle fraction changes  with
density. 

From Fig.  \ref{fig3c}
it is seen again that the $\alpha$-particle fraction is model dependent: 
GM1 predicts generally larger
fractions although the difference depends on the density.  The
differences can be as high as one or two orders of magnitude. The magnitude 
of $Y_\alpha$ depends on the fraction of protons present in the background gas. 
{ The low density onset of  the pasta-like matter in Figs. 5 and 6 lies
slightly below the homogeneous matter curves. The difference is smaller for 
the larger temperatures due to the distillation effect. The pasta-phase 
starts at low density with small droplets of dense matter in a
background gas. As soon as the droplets set on the  distillation
effect makes matter in the droplet more proton rich and in the gas
more proton poor. Therefore, fixing the global proton fraction, we
expect that at the onset of the droplets (the volume fraction of the dense part 
is very small)  the gas phase has a smaller proton fraction than the 
homogeneous matter at the corresponding density. This effect is
smaller for the larger temperatures and can be observed from
Figs. 5 and 6.}

\begin{table*}[thb]
\caption{Densities of the inner edge of the crust (crossing points)
 for NL3 }
\begin{center}
\begin{tabular}{llcccccc}
\hline
&EOS & $\rho_{t}^1~~/~~\rho_{t}^2$ & $P^1~~/~~P^2$ & pasta structure\\
 &   & fm$^{-3}$                      & MeV/fm$^3$    & \\
\hline
$T=5$ MeV\\
$Y_p=0.5$\\
 &no $\alpha$s (*) & 0.004 / 0.077 &  0.039  / 2.065 & droplets, rods, slabs \\
 & $\Gamma_{va}=0$ & 0.010 / 0.077 &  0.155 / 2.069  &  
droplets, rods, slabs \\
 & $\Gamma_{va}=4\Gamma_v$ & 0.004 / 0.077 & 0.036 / 2.065 &  
droplets, rods, slabs \\
$Y_p=0.3$\\
 &no $\alpha$s (*) & 0.003 / 0.082 & 0.015 / 1.136 & droplets, rods, slabs \\
 & $\Gamma_{va}=0$ & 0.006 / 0.082 &  0.044 / 1.138 &  
droplets, rods, slabs \\
 & $\Gamma_{va}=4\Gamma_v$ & 0.003 / 0.082 & 0.017 / 1.136 &  
droplets, rods, slabs \\
$Y_l=0.4$\\
 &no $\alpha$s &0.003 / 0.080& 0.024 / 1.430 &  droplets, rods,slabs\\
 & $\Gamma_{va}=0$ & 0.007 / 0.080& 0.060 / 1.431 &  droplets, rods,slabs\\
 & $\Gamma_{va}=4\Gamma_v$ &0.003 / 0.080& 0.024 / 1.430 &  droplets, rods,slabs
\\
\hline
$T=8$ MeV\\
$Y_p=0.3$\\
 &no $\alpha$s (*) & 0.025 / 0.042 & 0.275 / 0.516  & droplets, rods \\
 & $\Gamma_{va}=0$ & 0.025 / 0.045 & 0.281 / 0.572 &  
droplets, rods \\
 &$\Gamma_{va}=4\Gamma_v$ & 0.025 / 0.042 & 0.275 / 0.515  &  
droplets, rods \\
$Y_l=0.4$\\
 &no $\alpha$s  & -/-& -/-&\\
 & $\Gamma_{va}=0$ & 0.029 / 0.039&   0.418 / 0.593  & droplets, rods\\
 & $\Gamma_{va}=4\Gamma_v$  &-/-& -/-& \\
\hline
\end{tabular}
\end{center}
\label{tab2}
\end{table*}

The $\alpha$ particle densities are plotted in Fig. \ref{fig4} for two proton 
fractions ($Y_p=0.5$ and 0.3) and two temperatures ($T=5,\,8$ MeV). We include 
the  calculation for both  homogeneous matter and pasta-like matter and we 
only consider the case of interacting $\alpha$ particles. { This figure 
gives a hint on the possible effects of $\alpha$ particles in the inner crust
of a compact star.} The effect is larger for the larger temperatures
and larger proton fractions. Due to the existence of a non-homogeneous
phase the presence of the $\alpha$-particles extends to larger densities. 
{ In fact the 
$\alpha$ clusters dissolve at $\rho \sim$ 0.01 fm$^{-3}$ in homogeneous matter. 
If we 
consider the pasta phase the background gas does not exceed this value of the 
density and, therefore, there are still alpha-clusters at much larger 
densities because they only exist in the gas phase.}
However, we should point out, that with the strongly repulsive
interaction considered, $Y_\alpha$ takes very small values.

{\bf In Fig \ref{fig1}a) and \ref{fig2}a) it is clearly seen that the maximum
fraction of alpha particles decreases with temperature, however
clusters dissolve at larger densities and above 0.01 fm$^{-3}$ the
fraction of $\alpha$ particles is much larger for T=10 MeV than for T=5
MeV. This effect is still seen in the pasta phase:    above  0.01 fm$^{-3}$
$\alpha$ particle fractions are larger at $T=8$ MeV than at $T=5$ MeV.}

{ It is worth emphasing some points related to the effects of the
$\alpha$ particles present in the inner crust of a compact star. They are
small or negligible in calculations of the evolution of protoneutron stars
due to the very low densities involved. However, simulations of 
supernovae explosions seem to indicate the the internal structure of the pasta 
phase plays a decisive role in avoiding the stalling process. 
Moreover, at the  inner crust the shear viscosity, thermal conductivity and 
electrical 
conductivity are mainly determined by electron-ion scattering. Above neutron 
drip it is important also to consider electron-neutron scattering 
\cite{flowers76, chugonov05,horowitz08}.  It has been shown \cite{chamel08} 
that the shear  viscosity in the inner crust,  which is mainly determined by
electrons, is affected by the electron-impurity scattering at 
low temperatures. Recently, taking into account  electron-proton scattering it 
was shown in \cite{horowitz08} that the pasta structures would not increase as 
expected the shear viscosity and thermal conductivity due to an effective 
reduction of the proton number of the clusters owing to ion screening effects.
 We expect that the presence of alpha-particles will give an extra 
contribution that should be calculated. In this paper it was also shown that 
the scattering of neutrinos from neutrons would be defined by the difference 
between the neutron content of the pasta structures and the dripped neutron gas.
 The reduction of the background neutron gas due to the alpha particle 
formation could increase the effective neutron content of the pasta 
structures.}

In order to show the effect of the
$\alpha$ particle interaction on the background gas  of the pasta phase, we 
plot in  Fig. \ref{fig5}, for matter with trapped neutrinos
and a lepton fraction $Y_l=0.4$, 
the density  { $\rho_\alpha$} obtained with free and interacting  
particles.   The $\alpha$-particle density is lower for TW.  
A smaller $\alpha$-particle fraction is due to the smaller symmetry
energy for TW at subsaturation densities, which  
does not favor the formation of isospin symmetric clusters. GM1, on the other 
hand, has the largest symmetry energy and predicts much larger 
$\alpha$-particle fractions. 
 We also conclude that  knowing with accuracy the 
strength of the interaction is crucial to obtain the correct order 
of magnitude for  $Y_\alpha$.
 At $\rho=0.08$ fm$^{-3}$, which corresponds
approximately to the crust-core transition, the difference between a free $\alpha$ particle gas
and an interacting one with $\Gamma_\alpha=4\Gamma_v$ could be as high as 6 orders of magnitude.
 Reducing
the repulsive $\omega-\alpha$ interaction to one half,
$\Gamma_\alpha=2\Gamma_v$,  reduces the
difference to  3 orders of magnitude. This test is shown only for NL3 and is 
represented in
the figure by the red medium thick line, close to the GM1 result with  $\Gamma_\alpha=4\Gamma_v$.

\subsection{Effect of the $\alpha$ particles on the phase transitions}

  \begin{figure}
\begin{center}
\includegraphics[width=0.9\linewidth,angle=0]{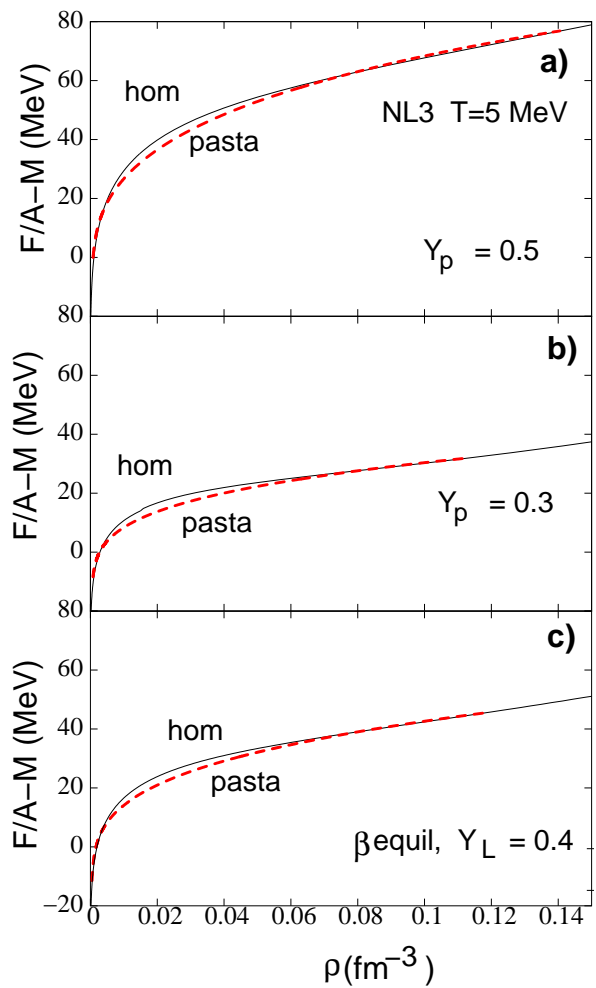} \\
\end{center}
\caption{(Color online) Free energy for the homogeneous and pasta like matter obtained within the NL3 for T=5 MeV and several proton fractions. The red dashed line defines the pasta free energy and its extension is determined by the binodal. The crossing of the pasta  and the homogeneous free energy define the extension of the pasta phase.}
\label{free}
\end{figure}

In Fig. \ref{free} we have plotted the free energy for homogeneous and pasta like matter described by the NL3 model with T=5 MeV and 
proton fractions $Y_p=0.5,0.3$ and for $\beta$-equilibrium matter with trapped 
neutrinos for a lepton fraction $Y_l=0.4$. The extension of the pasta free energy is determined by the binodal surface and becomes smaller when the proton fraction decreases. The system in equilibrium chooses the configuration with lower free energy so the pasta-like matter defines the ground state of the system 
only while its free energy is lower than the free energy of the corresponding 
homogeneous matter.

In Tables \ref{tab2} and \ref{tab2tw}  the densities at the crossing between 
the homogeneous and the pasta phases are given for NL3 and TW. The two
crossing phases correspond to homogeneous matter to pasta { ($\rho_t^1$ in 
the Tables)}, at very low densities, and pasta to homogeneous matter 
{ ($\rho_t^2$ in the Tables)}, at slightly higher densities. The 
corresponding pressures { ($P^1$ and $P^2$ obtained in the pasta phase)} 
are also given. The (*) means that the numbers shown for the crossing 
densities with NL3 and no $\alpha$ particles are different from the ones 
shown in \cite{pasta1} because a different parametrization for the surface 
energy was used here.

It is seen from Tables \ref{tab2} and \ref{tab2tw} that the transition density for symmetric matter is smaller than the  transition density for matter with $Y_p=0.3$ while the opposite would be expected since the binodal has a smaller extension for the smaller proton fractions. In fact a Thomas Fermi calculation predicts respectively 0.104 and 0.092 fm$^{-3}$ for $Y_p=0.5$ and $Y_p=0.3$ for NL3 and T= 5 MeV \cite{iwara09}. As can be seen from Fig. \ref{free} there is a range of densities close to the transition density when the free energy of pasta and homogeneous matter do not differ much. The same occurs with a Thomas Fermi calculation. This corresponds to the range of densities which is sensitive to the method used and the self-consistency of the method. One large drawback of the 
coexisting phases method is the fact the proton distribution inside the pasta structure is not perturbed by the Coulomb field. In particular, for symmetric matter the proton fraction of the structure is 0.5, larger than the prediction of a Thomas Fermi calculation which would predict that  protons spread giving rise to a  smaller  proton fraction, since the neutrons are not affected. This effect is larger the larger the proton fraction and is particularly critical for symmetric matter.  We have verified that the density transition for symmetric matter at T=5 MeV  would increase to 0.087 fm$^{-3}$ if the central proton fraction of the structure would reduce by 20\% due to the Coulomb force. 

\begin{table*}[thb]
\caption{Densities of the inner edge of the crust (crossing points)
 for TW}
\begin{center}
\begin{tabular}{llcccccc}
\hline
&EOS & $\rho_{t}^1~~ / ~~\rho_{t}^2$ & $P^1~~ / ~~P^2$ & pasta structure\\
 &   & (fm$^{-3})$                      & (MeV / fm$^3$)    & \\
\hline
$T=5$ MeV&\\
$Y_p=0.5$\\
&no $\alpha$s (*) & 0.004  /  0.079 &0.037   /  2.060  & droplets, rods, slabs \\
& $\Gamma_{va}=0$ & 0.012  /  0.079 &0.160   /  2.067&  
droplets, rods, slabs \\
& $\Gamma_{va}=4\Gamma_v$ & 0.004  /  0.079 &  0.038  / 2.060   &  
droplets, rods, slabs \\
$Y_p=0.3$\\
&no $\alpha$s (*) & 0.003  /  0.086 & 0.018  /  1.209 & droplets, rods, slabs \\
& $\Gamma_{va}=0$ & 0.007  /  0.086 & 0.049  / 1.211  &  
droplets, rods, slabs \\
& $\Gamma_{va}=4\Gamma_v$ & 0.003  /  0.086 & 0.018  / 1.209  &  
droplets, rods,slabs \\
$Y_l=0.4$\\
&no $\alpha$s (*) &  0.004  /  0.085 &0.028   /  1.536 & droplets, rods, slabs \\
& $\Gamma_{va}=0$ & 0.008  /  0.085 & 0.072  /  1.534 &  
droplets, rods, slabs \\
& $\Gamma_{va}=4\Gamma_v$ & 0.004  /   0.085 &  0.026 / 1.537  &  
droplets, rods,slabs \\
\hline
$T=8$ MeV&\\
$Y_p=0.5$\\
&no $\alpha$s (*) &  - /  - &  - /  - &  \\
& $\Gamma_{va}=0$ & 0.030  /  0.042 & 0.605  / 0.919 &  
droplets, rods \\
& $\Gamma_{va}=4\Gamma_v$ & -  /  - & -  / -  &  
 \\
$Y_p=0.3$\\
&no $\alpha$s (*) & 0.021  /  0.054 &0.221   /  0.714  & droplets, rods \\
& $\Gamma_{va}=0$ & 0.024  /  0.056 &0.263   /  0.747 &  
droplets, rods \\
&$\Gamma_{va}=4\Gamma_v$ & 0.021  /  0.054 &0.222   / 0.715  &  
droplets, rods \\
$Y_l=0.4$\\
&no $\alpha$s (*) &  0.023  /  0.050 & 0.306  /  0.816 & droplets, rods \\
& $\Gamma_{va}=0$ & 0.026  /  0.050 & 0.389  /  0.818  &  
droplets, rods\\
& $\Gamma_{va}=4\Gamma_v$ & 0.023  /   0.050 & 0.306   /  0.816 &  
droplets, rods \\
\hline
\end{tabular}
\end{center}
\label{tab2tw}
\end{table*}

\begin{table*}[h]
\caption{Densities of the inner edge of the crust (crossing points)
 for the T=5 MeV and $\Gamma_\alpha=4\Gamma_v$}
  \begin{tabular}{l c cc cc c c}
    \hline
 & & $\rho_{t}^1~~ / ~~\rho_{t}^2$& & $\phantom{mm}$& &$P^1~~ / ~~P^2$ &   \\
$\phantom{mmmmmmm}$ & & (fm$^{-3}$)  & & &  & (MeV fm$^{-3}$)&\\

 \cline{2-4} \cline{6-8}\\

&   NL3 & GM1 & TW & & NL3 & GM1 & TW\\
    \hline
$Y_p=0.5$ & 0.004 / 0.077 & 0.005 / 0.080 & 0.004 / 0.079& & 0.039  / 2.065 & 0.049 / 2.076 &   0.038  /  2.060\\ 
$Y_p=0.3$ & 0.003 / 0.082 & 0.004 / 0.088 & 0.003 / 0.086& & 0.017  / 1.136 &0.021  / 1.243& 0.018  /  1.209 \\
$Y_l=0.4$ & 0.003 / 0.080 & 0.003 / 0.094 & 0.004 / 0.085 && 0.024 / 1.430 & 0.028 / 1.527 & 0.026  /  1.537 \\
    \hline
  \end{tabular}
\label{tab3}
\end{table*}

The presence of the $\alpha$ particles has only a small effect on the pasta
phase structure { and on the size of the clusters}. The largest effect occurs for the lower densities, namely the low density limit of the
onset of the pasta phase, when a gas of free $\alpha$-particles is considered: this defines an
upper limit of the possible effect. The density at the  upper border may be slightly larger for
temperatures and proton fractions close to the critical values above which the pasta phase
disappears, like the results obtained for T=8 MeV. The density at the lower border is generally
larger due to the softer EOS for the homogeneous matter with $\alpha$-particles. Including the
interaction of the $\alpha$-particles with the $\omega$ meson { defines} 
the extension of the
pasta phase to the limits obtained without $\alpha$-particles. The effect of the model is
mainly noticeable  close the critical values of temperature and proton fraction: in particular
for matter with trapped neutrinos at T=8 MeV we predict a pasta phase within TW, with or
without $\alpha$-particles, however, for NL3  there is a pasta phase only in the presence
of non interacting $\alpha$ particles.
A more precise knowledge of the  effect of $\alpha$-particles requires a self-consistent
calculation, within, for instance a Thomas Fermi calculation.

We have included in Tables \ref{tab2} and \ref{tab2tw} the pressure at the transition between the pasta phase and the homogeneous matter because  it as been   shown in \cite{pethick-95} that  this  pressure  defines
the mass and moment of inertia of the crust of compact stars. The presence of $\alpha$ particles does not seem to have a large effect on the  pressure at the transition. 

{ At this point a comment on the effect of thermal fluctuations is in order.
The problem of the effect of  thermal fluctuations on the pasta
structures has been studied by \cite{thermal1,thermal2} and it was shown that 
thermally induced displacements of the rod-like and
slab-like nuclei could melt the lattice structure when these displacement are 
larger than the space available between the cluster and the boundary of the 
Wigner-Seitz cell. Using the elastic constants calculated in \cite{thermal1}
the authors of \cite{thermal2} have  calculated the critical
temperatures above which the ordered configuration is distroyed.
While for the rod like clusters and for the densities and temperatures
considered in our work the lattice would not be affected by the thermal
fluctuations, for the slabs and densities considered T=5 MeV is a
limiting temperature.}

Table \ref{tab3} allows a comparison between the models NL3, GM1 and TW at the crust-core transition for T=5 MeV and several proton fractions. The values in this table include the effect of interacting $\alpha$-particles. GM1 predicts the larger transition densities at the crust-core boundary. This reflects the larger binodal region this model presents.

\section{Conclusions}

In the present work we have studied the effect of $\alpha$ particles on warm 
low density
stellar matter as found in the inner edge of the crust of a protoneutron star.
We have considered three different types of relativistic nuclear models: 
the parametrization NL3 \cite{nl3} and GM1 \cite{glen}
of the non-linear Walecka model with constant couplings and the TW 
parametrization of the
density dependent relativistic hadronic model \cite{tw} with density
dependent coupling parameters. 

We have first considered an homogeneous  neutral gas  formed by protons, neutrons, electrons and
$\alpha$ particles with a fixed proton fraction or in $\beta$-equilibrium with trapped
neutrinos. The $\alpha$ particles were described as a gas of bosons and two opposite situations
were considered: a free gas of $\alpha$ particles with binding energy 28.3 MeV 
was used with no interaction  taken into
account, and a gas of particles with binding energy 28.3 MeV  interacting through the exchange of  $\omega$ meson which
reduces the interaction to a repulsive interaction. It is expected that a 
realistic situation lies between these two extrema. The binding energy is 
density and  temperature dependent and in the approach made in \cite{blaschke09} the inclusion
of a term that describes the temperature and density dependence of the binding energy is 
essential to dissolve the clusters.  In the present work this was not
considered explicitly, because, in order to reduce the unknown  $\alpha$ particle couplings,   the
attractive  $\sigma$ meson-$\alpha$ interaction  was not included.
 It was shown that the 
inclusion of the repulsive
interaction is essential to avoid an overprediction of $\alpha$ particles above
$\rho\sim0.001$ fm$^{-3}$ and it is also the mechanism responsible for the 
dissolution of the $\alpha$
particle clusters in the present approach. In the low density limit ($\rho<0.0001$fm$^{-3}$) all models considered behave in a similar way: this behavior essentially  coincides with the one predicted by the virial expansion
\cite{hor06} when the interparticle interaction is negligible, and the system behavior is model independent. The
main differences between models occur for $ \rho>0.001$ fm$^{-3}$ when 
the $\alpha$ particle fractions differences between models may be as large as 
one order of magnitude or even larger. TW predicts the smallest
fractions while GM1 the largest ones. Temperature shifts the maximum on 
the $\alpha$ particle
distribution and the density of cluster dissolution to larger densities, 
although the maximum
values of the distributions themselves decrease with
temperature. The maximum values of the $\alpha$ particle distributions also 
decrease when the
proton fraction decreases. However, the proton fraction has no effect on the 
density
localization of the maximum and neither on the density of dissolution of 
the clusters.  
 
We have next investigated the pasta phase with $\alpha$ particles. We have described the pasta
phase using the method described in \cite{maruyama,pasta1,pasta2}: the coexisting phases are
determined from Gibbs conditions and surface energy and Coulomb interaction are added {\em a
posteriori}. For the surface energy we have used the parametrizations determined  in terms of the proton fraction
and the temperature using a Thomas-Fermi calculation. It was shown in
\cite{maruyama,pasta1,pasta2} that the method of the coexisting phases is very sensitive to
the surface energy. We have performed our calculations for two temperatures $T=5$ and 8 MeV,
two proton fraction $Y_p=0.5$ and 0.3 and for $\beta$-equilibrium matter with trapped
neutrinos and a lepton fraction of 0.4. We have analyzed the fraction of $\alpha$
particles as a function of density in the pasta phase and,  comparing with the
fraction of $\alpha$ particles in homogeneous matter, it was shown that it was larger by many orders of magnitude
for densities above 0.01 fm$^{-3}$. It is important to stress that the
prediction obtained within a homogeneous EOS calculation underestimates the $\alpha$ particle fraction. This certainly affects transport properties such as
heat conduction and electrical conductivity. 

It was also seen that while for symmetric matter
the  $\alpha$ particle fraction decreases with temperature when interacting $\alpha$ particles
are considered, for $Y_p=0.3$ the opposite occurs. This is an interesting effect related to
 the proton fraction in the  background gas  which increases with temperature for asymmetric matter.

Finally we have analyzed the effect of the $\alpha$ particles on the pasta structure. It was
shown that the effect is 
small except close to the critical temperatures and / or proton
fractions when it may still predict a pasta phase while no pasta phase would occur in the
absence of light clusters. A self-consistent calculation is necessary to give more quantitative
predictions.  Other small clusters should also be included in the calculation.

\section{Appendix}

The surface tension coefficient, $\sigma$, was calculated in the Thomas-Fermi approximation and fitted to the functional form given in, 
eq.(\ref{sigpar}), with $\sigma(x)$=$\sigma(x,T=0)$, $a(T)$ and 
$b(T)$ having the following expressions:
\begin{eqnarray}
\sigma(x)= \sigma_0 \exp(-\sigma_1~x^{\frac{3}{2}})  (1+a_1~x+a_2~x^2+ \nonumber \\    
a_3~x^3+a_4~x^4+a_5~x^5+ a_6~x^6) ~~, \nonumber
\end{eqnarray}
$$
a(T)=a_0+a_1~T+a_2~T^2+a_3~T^3 ~~,
$$
$$
b(T)=a_0+a_1~T+a_2~T^2+a_3~T^3 ~~.
$$
The results of the model fitting using the Thomas-Fermi approximation for the GM1, NL3 and TW parametrizations are
given in table \ref{tabsig} for temperatures up to $T=10$ MeV. 

\begin{table}[h]
\caption{Surface tension coefficient parameters fitted within the Thomas-Fermi approximation for GM1\cite{glen}, NL3\cite{nl3} and TW\cite{tw} parametrizations.}
\label{tabsig}
\begin{center}
\begin{tabular}{cccc}
\hline
 GM1 & $\sigma(x)$ & $a(T)$ & $b(T)$  \\
\hline
  $\sigma_0$ & 1.48801 & - & -  \\
  $\sigma_1$ & 10.6647 & - & -  \\
  $a_0$ & - & 0.00521594 & 0.00640879  \\  
  $a_1$ & -4.32419 &0.0125187 & 0.000140886  \\  
  $a_2$ & 61.6172 & -0.00156999& -4.17615e-5  \\  
  $a_3$ & -488.383 &  8.4792e-5& 1.6856e-6  \\  
  $a_4$ & 1678.95 & - & -  \\  
  $a_5$ & -2450.41 & - & -  \\  
  $a_6$ & 1296.86 & - & -  \\ 
\hline
 NL3 & $\sigma(x)$ & $a(T)$ & $b(T)$  \\
\hline
  $\sigma_0$ & 1.12307 & - & -  \\
  $\sigma_1$ & 20.7779 & - & -  \\
  $a_0$ & - & 0.0121222 & 0.00792168  \\  
  $a_1$ & -5.84915 & 0.01664& -8.2504e-5  \\  
  $a_2$ & 138.839 & -0.00137266 & -4.59336e-6  \\  
  $a_3$ & -1631.42 & 4.0257e-5 & -2.81679e-7  \\  
  $a_4$ & 8900.34 & - & -  \\  
  $a_5$ & -21592.3 & - & -  \\  
  $a_6$ & 20858.6 & - & -  \\ \hline
 TW & $\sigma(x)$ & $a(T)$ & $b(T)$  \\
\hline
  $\sigma_0$ & 1.21736 & - & -  \\
  $\sigma_1$ & 8.33425 & - & -  \\
  $a_0$ & - & -0.00481823 & 0.00844664  \\  
  $a_1$ & -2.18075 & 0.0066498 & -7.23379e-4  \\  
  $a_2$ & 18.0584 & -0.000267288 & 9.6817e-5  \\  
  $a_3$ & -96.548 & 1.34544e-5 & -5.08488e-6  \\  
  $a_4$ & 259.517 & - & -  \\  
  $a_5$ & -296.69 & - & -  \\  
  $a_6$ & 125.94 & - & -  \\   
\hline   
\end{tabular}
\end{center}
\end{table}

\section*{ACKNOWLEDGMENTS}
This work was partially supported by Capes / FCT n. 232 / 09 bilateral 
collaboration, by CNPq (Brazil), by FCT and FEDER (Portugal) under the project
CERN / FP / 109316 / 2009 and  by Compstar, an ESF Research Networking Programme.


\begin{thebibliography}{99}

\bibitem{pethick} D. Ravenhall, C.J. Pethick and J.R. Wilson,
  Phys. Rev. Lett. {\bf 50}, 2066 (1983).

\bibitem{hashimoto}  M. Hashimoto, H. Seki and M. Yamada, Prog. Theor. Phys. {\bf 71}, 
320 (1984).

\bibitem{horo} C.J. Horowitz, M.A. P\'erez-Garcia and J. Piekarewicz,
  Phys. Rev. {\bf C 69}, 045804 (2004); C.J. Horowitz, M.A. P\'erez-Garcia,
D.K. Berry and J. Piekarewicz, Phys. Rev. {\bf C 72}, 035801 (2006).

\bibitem{watanabe} G. Watanabe, K. Sato, K. Yasuoka, and T. Ebisuzaki.
Phys. Rev. C 66, 012801 (2002);
G. Watanabe, K. Sato, K. Yasuoka, and T. Ebisuzaki, Phys. Rev. C 68, 
035806 (2003);
G. Watanabe, K. Sato, K. Yasuoka, and T. Ebisuzaki, Phys. Rev. C 69, 055805 
(2004);
H. Sonoda, G. Watanabe, K. Sato, K. Yasuoka, and T. Ebisuzaki,
Phys. Rev. C 77, 035806 (2008).

\bibitem{maruyama} T. Maruyama, T. Tatsumi, D.N. Voskresensky, T. Tanigawa and
  S. Chiba,  Phys. Rev. {\bf C 72}, 015802 (2005).

\bibitem{pasta1} S.S. Avancini, D.P. Menezes, M.D. Alloy, J.R. Marinelli, 
M.M.W. de Moraes and C. Provid\^encia,  Phys. Rev. {\bf C 78}, 015802 (2008). 

\bibitem{bao} J. Xu, L.W. Chen, B.A. Li and H.R. Ma, arXiv:0807.4477v1 
[nucl-th].

\bibitem{pasta2} S.S. Avancini, L. Brito, J.R. Marinelli, D.P. Menezes, 
M.M.W. de Moraes, C. Provid\^encia and A.M. Santos - Phys. Rev. {\bf C 79}, 
035804 (2009).

\bibitem{watanabe05}G. Watanabe and H. Sonoda, cond-mat / 0502515.


\bibitem{sw} B. Serot and J.D. Walecka, {\em Advances in Nuclear Physics} 16, Plenum-Press, (1986) 1.

\bibitem{nl3} G. A. Lalazissis, J. K\"onig and P. Ring, Phys. Rev. {\bf C 55}, 540 (1997).

\bibitem{tm1} K. Sumiyoshi, H. Kuwabara, H. Toki, Nucl. Phys. {\bf A581}, 725 (1995).

\bibitem{glen} N. K. Glendenning, Compact Stars, Springer-Verlag, New-York,
2000.

\bibitem{tw} S. Typel and H. H. Wolter, Nucl. Phys. {\bf A656}, 331 (1999);
G. Hua, L.Bo and M. Di Toro, Phys. Rev. {\bf C  62}, 035203(2000).

\bibitem{gaitanos} T. Gaitanos, M. Di Toro, S. Typel, V. Baran, C. Fuchs,
V. Greco  and H. H. Wolter, Nucl. Phys. {\bf A 732}, 24 (2004).

\bibitem{gogelein}P. G\"ogelein, E.N.E. van Dalen, C. Fuchs and H. M\"uther, 
Phys. Rev. {\bf C 77}, 025802 (2008).

\bibitem{dalen}E.N.E. van Dalen, C. Fuchs and A. Faessler, Eur. Phys. J.A. {\bf 31}, 29(2007).

\bibitem{camille08} C. Ducoin, C. Provid\^encia, A. M. Santos, L. Brito and 
Ph. Chomaz, Phys. Rev. {\bf C 78}, 055801 (2008).

\bibitem{watanabe09} Gentaro Watanabe, Hidetaka Sonoda, Toshiki Maruyama, Katsuhiko Sato, Kenji Yasuoka, Toshikazu Ebisuzaki, Phys. Rev. Lett. 103, 121101 (2009) 

\bibitem{ls91} J. M. Lattimer and  F. D. Swesty, Nucl. Phys. A 535, 331 (1991)
\bibitem{shen} H. Shen, H. Toki, K. Oyamatsu, K. Sumiyoshi, Nucl. Phys. 
{\bf A 637}, 435 (1998).

\bibitem{roepke} M. Beyer, S.A. Sofianos, C. Kuhrts, G. Roepke and P. Schuck,
arXiv:nucl-th/0003071v4.

\bibitem{hor06} C.J. Horowitz and A. Schwenk,  Nucl. Phys. A 776, 55 (2006).

\bibitem{blaschke09} S. Typel, G. Roepke, T. Klahn, D. Blaschke and H.H. Wolter,
Phys. Rev. C 81, 015803 (2010).

\bibitem{shenguides} H. Shen, H. Toki, K. Oyamatsu and K. Sumiyoshi, {\it User 
Notes for Relativistic EOS Table}, private communication.


\bibitem{nse} W.R. Hix, F.K. Thielemann, I. Fushiki, J.W. Truran, {\it Nuclear
Statistical Equilibrium and the Influence of Coulomb Screening - chapter 1},
private communication.

\bibitem{roepke03} G. Roepke, A. Grigo, K. Sumiyoshi and H. Shen, 
Nato Science Series, Series II: Mathematics, Physics and Chemistry 197, 75 
(2006). 

\bibitem{cphelena} H. Pais, A. Santos and C. Provid\^encia, Phys. Rev. C 
{\bf 80}, 045808 (2009). 


\bibitem{inst04} S.S. Avancini, L. Brito, D. P. Menezes and C. Provid\^encia,
Phys. Rev. C {\bf 70}, 015203 (2004).

\bibitem{watanabe2001} G. Watanabe, K. Iida and K. Sato, Nucl. Phys. A676, 455  (2000);
G. Watanabe, K. Iida and K. Sato, Nucl. Phys. A687, 512  (2001); G. Watanabe, K. Iida and
K. Sato, Nucl. Phys. A726, 357 (2003)

\bibitem{lattimer} J.M. Lattimer, C.J. Pethick, D.G. Ravenhall, D.Q. Lamb, 
Nucl. Phys. {\bf A 432}, 646 (1985).  

\bibitem{centel} M. Centelles, M. Del Estal and X. Vi\~nas, 
Nucl. Phys. {\bf A 635}, 193 (1998).  

\bibitem{mayers} W. D. Mayers and W. J. Swiatecki,
Phys. Rev. C {\bf 63}, 034318 (2001).

\bibitem{marina} M. Nielsen and J. da Provid\^encia, J. Phys. G:
  Nucl. Part. Phys. {\bf 16}, 649 (1990). 


\bibitem{cpsig} C. Provid\^encia, L. Brito and J. da Provid\^encia,
M. Nielsen and X. Vin\~nas, Phys. Rev. C {\bf 54}, 2525 (1996).

\bibitem{dmcp} D. P. Menezes and C. Provid\^encia,
Phys. Rev. C {\bf 60}, 024313 (1999).

\bibitem{flowers76} Elliott Flowers and Naoki Itoh, Astrophys. J. 206, 218 
(1976).

\bibitem{chugonov05} A. I. Chugonov and D. G. Yakovlev, Astronomy Reports 49, 
814 (2005).

\bibitem{horowitz08} C. J. Horowitz and D. K. Berry, Phys. Rev C 78, 035806 
(2008).

\bibitem{chamel08}Nicolas Chamel and Pawel Haensel,
Living Rev. Relativity 11, 10  (2008);  URL
http://www.livingreviews.org/lrr-2008-10

\bibitem{iwara09} S. S. Avancini, J. R. Marinelli, D. P. Menezes, 
M. M. W. Moraes, C. Provid\^encia, A. M. Santos, 
Int. J. Mod. Phys. D, accepted. 

\bibitem{inst062} S.S. Avancini, L. Brito, Ph. Chomaz, D. P. Menezes and C.
Provid\^encia, Phys. Rev. C {\bf 74}, 024317 (2006).

\bibitem{pethick-95}%
C. J. Pethick and D.~G.
Ravenhall, Ann.\ Rev.\ Nucl.\ Part. Sci. 45, 429 (1995)

\bibitem{thermal1} C. J. Pethick and A. Y. Potekhin, 
Phys. Lett. B 427, 7 (1997)

\bibitem{thermal2} G. Watanabe, K. Iida and K. Sato, Nucl. Phys. A 676, 455 
(2000), Nucl. Phys. A 726 (2003) 357



\end{thebibliography}
\end{document}